 [2005/06/22 5.2/AAS markup document class]%

\documentclass[preprint2]{aastex}%

\makeindex
\makeatletter
\let\o@verbatim\verbatim
\def\verbatim{%
  \ifhmode\unskip\par\fi
  \ifx\@currsize\normalsize
     \small
  \fi
  \o@verbatim
}

\renewcommand \verbatim@font {%
  \normalfont \ttfamily
  \catcode`\<=\active
  \catcode`\>=\active
}

\RequirePackage{shortvrb}
\MakeShortVerb{\|}

\begingroup
  \catcode`\<=\active
  \catcode`\>=\active
  \gdef<{\@ifnextchar<\@lt\@meta}
  \gdef>{\@ifnextchar>\@gt\@gtr@err}
  \gdef\@meta#1>{\m{#1}}
  \gdef\@lt<{\char`\<}
  \gdef\@gt>{\char`\>}
\endgroup
\def\@gtr@err{%
   \ClassError{ltxguide}{%
      Isolated \protect>%
   }{%
      In this document class, \protect<...\protect>
      is used to indicate a parameter.\MessageBreak
      I've just found a \protect> on its own.
      Perhaps you meant to type \protect>\protect>?
   }%
}
\def\verbatim@nolig@list{\do\`\do\,\do\'\do\-}

\newcommand{\m}[1]{\mbox{\it #1}}

\def\cmd#1{\cs{\expandafter\cmd@to@cs\string#1}}
\def\cmd@to@cs#1#2{\char\number`#2\relax}
\DeclareRobustCommand\cs[1]{\texttt{\char`\\#1}}
 \usepackage{amsmath}    
 \usepackage{graphicx}   
 \usepackage{verbatim}   
 \usepackage{xspace}

 \usepackage{amssymb}

 \usepackage{lineno}

 \usepackage[caption=false]{subfig}
 \DeclareCaptionListOfFormat{myparens}{#2}
 \newcommand{\Subref}[1]{\protect\subref{#1}}

 \usepackage[dvipsnames,usenames]{color}
 \usepackage[breaklinks={true}, bookmarks, colorlinks={true}, pdfpagemode={NON}, pdffitwindow=true, pdfstartview={FitB}, linkcolor={BlueViolet}, menucolor={BlueViolet}, citecolor={Mahogany}, filecolor={OliveGreen}, urlcolor={BlueViolet}, pdftitle={ANNz2 - photometric redshift and probability distribution function estimation using machine learning}, pdfauthor={Iftach Sadeh}, pdfsubject={Photometric redshifts and probability distribution functions}, pdfkeywords={Photometric redshifts; Machine learning}]{hyperref}

 \let\orgautoref\autoref
 \providecommand{\Autoref}
         {\def\equationautorefname{Equation}%
          \def\figureautorefname{Figure}%
          \def\subfigureautorefname{Figure}%
          \def\chapterautorefname{Chapter}%
          \def\sectionautorefname{Section}%
          \def\subsectionautorefname{Section}%
          \def\subsubsectionautorefname{Section}%
          \def\Itemautorefname{Item}%
          \def\tableautorefname{Table}%
          \def\appendixautorefname{Appendix}%
          \orgautoref}

 \providecommand{\Autorefs}
         {\def\equationautorefname{Equations}%
          \def\figureautorefname{Figures}%
          \def\subfigureautorefname{Figures}%
          \def\chapterautorefname{Chapters}%
          \def\sectionautorefname{Sections}%
          \def\subsectionautorefname{Sections}%
          \def\subsubsectionautorefname{Sections}%
          \def\Itemautorefname{Items}%
          \def\tableautorefname{Tables}%
          \def\appendixautorefname{Appendices}%
          \orgautoref}

 \renewcommand{\autoref}
         {\def\equationautorefname{Eq.}%
          \def\figureautorefname{Fig.}%
          \def\subfigureautorefname{Fig.}%
          \def\chapterautorefname{Ch.}%
          \def\sectionautorefname{Sect.}%
          \def\subsectionautorefname{Sect.}%
          \def\subsubsectionautorefname{Sect.}%
          \def\Itemautorefname{item}%
          \def\tableautorefname{Table}%
          \orgautoref}

 \newcommand{\headFont}[1]{\textbf{\bfseries#1}}

 \newcommand{\eg}        {e.g.\/,\xspace}

 \newcommand{\mrm}[1]{\mathrm{#1}}
 \newcommand{\ttt}[1]{\texttt{#1}}

 \newcommand{\sixEightP}{\ensuremath{68^{\textrm{th}}~\textrm{percentile}}\xspace}

 \newcommand{\annzOne}{\texttt{ANNz1}\xspace}
 \newcommand{\annz}{\texttt{ANNz2}\xspace}
 \newcommand{\ROOT}{\texttt{ROOT}\xspace}
 \newcommand{\tmva}{\texttt{TMVA}\xspace}
 \newcommand{\skynet}{\texttt{SkyNet}\xspace}
 \newcommand{\arborz}{\texttt{ArborZ}\xspace}
 \newcommand{\tpz}{\texttt{TPZ}\xspace}
 \newcommand{\lephare}{\texttt{Le\kern -0.3em~PHARE}\xspace}
 \newcommand{\bpz}{\texttt{BPZ}\xspace}
 \newcommand{\zebra}{\texttt{ZEBRA}\xspace}

 \newcommand{\Photoz}{Photo-\ensuremath{z}\xspace}
 \newcommand{\photoz}{photo-\ensuremath{z}\xspace}
 \newcommand{\Photozs}{Photo-\ensuremath{z}s\xspace}
 \newcommand{\photozs}{photo-\ensuremath{z}s\xspace}

 \newcommand{\kdTree}{kd-tree\xspace}

 \newcommand{\zSpec}{\ensuremath{z_{\mathrm{spec}}}\xspace}
 \newcommand{\zPhot}{\ensuremath{z_{\mathrm{phot}}}\xspace}
 \newcommand{\zBest}{\ensuremath{z_{\mathrm{best}}}\xspace}
 \newcommand{\qNN}{\ensuremath{Q_{\mathrm{NN}}}\xspace}

 \newcommand{\PDF}{\ensuremath{\mathrm{\bf{PDF}}}\xspace}

 \newcommand{\Ssb}{\ensuremath{S_{\mathrm{s/\ \kern -0.2em b}}}\xspace}

 \usepackage{listings}
 \definecolor{listings_whitesmoke}{rgb}{0.96, 0.96, 0.96}
 \definecolor{listings_prussianblue}{rgb}{0.0, 0.19, 0.33}
 \definecolor{listings_cerise}{rgb}{0.87, 0.19, 0.39}
 \definecolor{listings_mediumpersianblue}{rgb}{0.0, 0.4, 0.65}
 \definecolor{listings_sapgreen}{rgb}{0.31, 0.49, 0.16}
 \definecolor{listings_olive}{rgb}{0.42, 0.56, 0.14}

\makeatother
\title{%
ANNz2 - photometric redshift and probability\\ distribution function estimation using machine learning
}%

\author{I.~Sadeh,$^{\ast\dagger}$ F.~B.~Abdalla$^{\ast\ddagger}$ and O.~Lahav$^{\ast}$\vspace{10pt}}
\affil{$^{\ast}$~Department of Physics \& Astronomy, University College London, Gower Street, London WC1E 6BT, UK}
\affil{$^{\dagger}$~DESY-Zeuthen, D-15738 Zeuthen, Germany}
\affil{$^{\ddagger}$~Department of Physics and Electronics, Rhodes University, PO Box 94, Grahamstown, 6140, South Africa}

\shorttitle{ANNz2 - Photometric redshift and PDF estimation}
\shortauthors{I.~Sadeh, F.~B.~Abdalla and O.~Lahav}

\begin{document}

\begin{abstract}
We present \annz, a new implementation of the public software for photometric
redshift (\photoz) estimation of~\cite{Collister:2003cz}, which now includes generation
of full probability distribution functions (PDFs).
\annz utilizes multiple machine learning methods, such as artificial neural networks and boosted
decision/regression trees.
The objective of the algorithm is to 
optimize the performance of the \photoz estimation,
to properly derive the associated uncertainties, and to produce both single-value solutions and PDFs.
In addition, estimators are made available, which mitigate possible problems of
non-representative or incomplete spectroscopic training samples.
\annz has already been used as part of the first weak lensing analysis of the
Dark Energy Survey, and is included in the experiment's first public data release.
Here we illustrate the functionality of the code using data from
the tenth data release of the Sloan Digital Sky Survey and the Baryon Oscillation Spectroscopic Survey.
The code is available for download at \href{https://github.com/IftachSadeh/ANNZ}{https://github.com/IftachSadeh/ANNZ}~.
\end{abstract}



\maketitle

\section{Introduction}\label{SECintroduction}
%
\subsection{Photometric redshifts}
%
Redshifts, usually denoted by~$z$, effectively provide a third, radial dimension to Cosmological analyses.
They allow the study of phenomena as a function of distance and time, as well as enable the identification of large structures, such as galaxy clusters. 
The current and next generations of dark energy experiments, such as
the Dark Energy Survey (DES),\footnote{~See \href{http://www.darkenergysurvey.org}{http://www.darkenergysurvey.org}~.}
the Large Synoptic Survey Telescope
(LSST),\footnote{~See \href{http://www.lsst.org}{http://www.lsst.org}~.}
and the Euclid experiment\footnote{~See \href{http://sci.esa.int/euclid/}{http://sci.esa.int/euclid/}~.}
will collectively observe a few billion galaxies.
Ideally, redshifts may be measured with great precision using spectroscopy. However, 
it is infeasible to obtain spectra for such large galaxy samples.
The success of these imaging surveys is therefore
critically dependent on the measurement of high-quality photometric redshifts (\photozs).
For instance, a benchmark of LSST is to measure the dark energy
equation of state parameter, $w$, with per-cent level uncertainty. This is
expected to be achievable with weak lensing tomography~\citep{Hu:1999ek,Zhan:2006jt}.
However, it will require a precision of~${\sim0.002 \cdot \left( 1 + z \right)}$ in determination of the systematic
bias in the redshift.

This paper presents \annz. The latter is a new implementation of the code of~\citet{Collister:2003cz},
denoted hereafter as \annzOne,
which used artificial neural networks to estimate photometric redshifts.
\annz is free and publicly
available.\footnote{~See \href{https://github.com/IftachSadeh/ANNZ}{https://github.com/IftachSadeh/ANNZ}~.}
The code has already been incorporated
as part of the analysis chain of DES~\citep{2014MNRAS.445.1482S}. It has been shown
to provide reliable \photoz estimates and to reduce systematic uncertainties and outlier
contamination~\citep{2015arXiv150705647L}. \annz
\photozs were part of the first DES weak lensing analysis~\citep{2015arXiv150705552T,2015arXiv150705909B},
are included in the first public data
release of the project,\footnote{~See
\href{http://des.ncsa.illinois.edu/releases/sva1}{http://des.ncsa.illinois.edu/releases/sva1}~.}
and are being used for upcoming analyses.

The extensive work in the community on \photozs usually falls into two categories,
papers on particular methods (see below)
and studies comparing existing methods~\citep{Abdalla:2008ze,2014MNRAS.445.1482S}.
The new ingredient of the present paper is a new approach to contrast and
combine different machine learning techniques, 
and to yield self-consistently a \photoz probability distribution function (PDF).
The introduction of PDFs has been shown to improve the accuracy of Cosmological
measurements~\citep{Mandelbaum:2007dp,Myers.2009}, and is an important new feature compared
to the previous version of the code.
In addition to \photoz inference, it is also possible to run \annz in classification
mode. The latter is useful for analyses
such as star/galaxy separation and morphological classification of galaxies.
An example is provided as part of the software package, but is not discussed in the following.

In the next section we present a short overview of the current methodology for deriving photometric
redshifts, focusing on machine learning, and on the techniques available through \annz.
We then describe the main methods implemented in the code for estimating \photozs and
PDFs, and illustrate the performance using a toy analysis. A short quick-start
guide for using the code is presented in the appendix.

\subsection{Methodology for \photoz estimation}\label{SECmethodologyForPhotometricEstimation}
%
The different approaches to calculate \photozs can generally be divided into two categories, template fitting methods and
training based machine learning. Both types depend heavily on photometric information, such as the integrated flux of photons in 
medium- or broad-band filters, which are usually converted into magnitudes or colours.
The magnitudes serve as a rough measurement of the underlying spectral energy distribution (SED)
of a target object, from which the redshift may be inferred.
A review of current \photoz methods can be found in~\citet{Abdalla:2008ze,Hildebrandt.2010}.
All methods require a spectroscopic dataset for training and/or calibration, the requirements for which
are discussed by~\citet{Newman:2013cac}.

Template fitting methods involve fitting empirical or synthetic galaxy spectra with
the photometric observables of an imaging survey, accounting for the response of the telescope and
the properties of the filters~\citep{Benitez:2008ts,Mobasher:2006fj}.
The template spectra are generally derived from a small set of SEDs, representing different classes of galaxies at zero redshift.
They also incorporate astrophysical effects, such as dust extinction in the Milky Way or in the observed galaxy.
Common template libraries are the~\citet{ColemanWuWeedman.1980} SEDs (derived observationally),
or those of~\citet{Bruzual.A.:1993is} (based on synthetic models). 

Template methods rely on the assumption that the SED templates are a true
representation of the observed SEDs. They depend \eg on proper calibration of the rest-frame spectra of galaxies,
commonly performed using spectroscopic data.
In addition, the composition of the template library should correspond to the population
of galaxies which are fitted (for instance, in terms of galaxy types and luminosities).
\Photozs may be estimated by choosing the best-fitted SED from the template library, usually derived using
$\chi^{2}$~-~minimization~\citep{Bolzonella:2000js},
where more advanced Bayesian priors can also be incorporated~\citep{Benitez:1998br}.

On the other hand, empirical methods do not directly use physically motivated models. Instead, they
involve deriving the relationship between the
photometric observables and the redshift using a so-called training dataset, which includes
both the observables and precise redshift information.
The mapping between observables and the output redshift can be as simple as a polynomial fit~\citep{Connolly:1995yq}.
However, supervised machine learning methods (defined below)\footnote{~Un-supervised learning
techniques have been used to derive photometric redshifts as well
(see \eg~\citet{2012MNRAS.419.2633G,2012PASP..124..274W,2014MNRAS.438.3409C}), but are not discussed here.}
have been shown to produce much more accurate
and robust results, taking into account complicated correlations between the input parameters and the
output value.

Machine learning methods have several advantages over template fitters.
For instance, it is trivial to incorporate additional observables into the inference, a common example being
the surface brightness of galaxies, which has a~${\left( 1 + z \right)^{-4}}$ redshift dependence~\citep{Firth:2002yz}.
In addition, the use of a training sample
alleviates systematic side-effects associated with the photometry, such as errors in the
zero-point corrections of the magnitudes.
On the other hand, the size and composition of the training sample become important factors in the performance.
The phase space of input parameters and the spectral types of galaxies
must correspond to the respective parameters of the survey. If this is not the case,
the \photozs of certain galaxy populations may become biased~\citep{2015MNRAS.450..305H}.
Another important point, is that the true redshift distribution in the spectroscopic training set must also
be representative of the survey. In particular, machine learning methods are only
reliable within the redshift range of the galaxies used for the training.
Consequently, they should not be used to
infer the \photozs of very high-redshift sources, for which there are no spectroscopic training data.
In order to resolve these problems, it is possible to generate synthetic training galaxies
within the required parameter space, using template-SED libraries. However, this introduces some of the
systematic biases associated with template fitting methods.

An important element of any \photoz algorithm is calculation of the associated uncertainty.
Accurate \photoz uncertainties help to identify catastrophic outliers, the removal of which
may improve the quality of Cosmological analyses~\citep{Abdalla:2007uc,Banerji:2007nh}.
For the previous version of the code, \annzOne, uncertainties were derived using a chain rule,
propagating the uncertainties on the algorithm-inputs (\eg magnitudes)
to an uncertainty on the final value of the \photoz.
Other methods exist~\citep{Oyaizu:2007jw},
which use the training data and the \photozs
themselves for uncertainty estimation. In these cases, the uncertainty is parametrized as a function of the inputs to the algorithm,
requiring no measurement of the uncertainties on the individual inputs.
We use such a scheme in \annz (see \autoref{SECsingleRegression}).

The common method for deriving the uncertainties for template fitting methods is
by combining the likelihoods estimated for the various templates.
The benefit of performing the combination is that it naturally leads to the definition of a \photoz PDF,
as, for instance, is the case in \lephare~\citep{Arnouts:1999bb,Ilbert:2006dp},
\bpz~\citep{Benitez:1998br} and \zebra~\citep{Feldmann:2006wg}.
As for machine learning methods, there is a variety of codes on the market.
These use different methods besides artificial neural networks,
such as boosted decision trees.
While most algorithms produce only single-value \photozs, several also generate
\photoz PDFs, such as \arborz~\citep{Gerdes:2009tw},
\tpz~\citep{Kind:2013eka}, \skynet~\citep{Bonnett:2013bza} and the algorithm of~\citet{Rau:2015sha}.
In \annz, two primary types of PDF are derived, one of which represents a new technique,
while the other is similar in nature to the PDFs generated by \arborz, \tpz and \skynet.

In the following, we describe in more detail the general workings of machine learning,
focusing on the primary algorithms used in \annz.

\subsection{Machine learning methods}
%
\subsubsection{Basics of machine learning}
%
Machine learning methods (MLMs) use supervised learning,
a machine learning task of inferring a function from a set of training examples.
Each example consists of an input object, described by a collection of
\textit{input parameters}, as well as a desired \textit{output value} for the MLM.
The training examples are used
to determine the mapping for either \textit{classification} or \textit{regression}
problems. The former describes a decision boundary between \textit{signal} and
\textit{background} entries; the latter refers to
an approximation of the underlying functional behaviour defining the output.

For the purpose of creating an MLM estimator for either classification or regression,
one generally splits the available dataset of examples into three parts, designated as the \textit{training},
\textit{validation} and \textit{testing} samples.
The training dataset is used for deriving the desired mapping between the input
and the output. During each step of the training, the validation sample is used to
estimate the convergence of the solution, by comparing
the result of the estimator with the value of the output. The testing dataset
is not used during the training process; rather, it is utilized
as an independent test of the performance of the trained MLM.

The MLMs utilized in \annz are implemented in the \tmva
package\footnote{~See
\href{http://tmva.sourceforge.net}{http://tmva.sourceforge.net}~.}~\citep{Hocker:2007ht}, which
is part of the \ROOT\ttt{C$++$} software
framework\footnote{~See
\href{http://root.cern.ch}{http://root.cern.ch}~.}~\citep{Brun:1997pa}. \tmva includes multiple MLM
methods, all of which are available through \annz, using a common \ttt{Python}
interface with simple control-options (see~\autoref{SECappendix}).
The two \tmva MLMs which we found to be most
appropriate for the problem of \photoz estimation, are artificial neural networks and boosted decision/regression trees.
For completeness, these are outlined concisely in the following. Detailed descriptions
of the implementation may be found in the \tmva manual.
For a comprehensive theoretical overview, see~\citet{mackay2003information,hastie01statisticallearning}.

\subsubsection{Available methods in \tmva}
%
\vspace{5pt}\noindent\headFont{-~Artificial neural networks (ANNs).}
One may consider an ANN as a mapping between a set of input variables, such as magnitudes or colours,
and one or more output variables. For regression problems, the output is \eg the numerical
value of a photometric redshift. For classification, the output is a variable (usually between~$0$ and~$1$),
which may be used to discriminate between signal and background examples.
The mapping is performed by computing the weighted sum of a collection of response functions. The
input variables, response functions and output variables are collectively called \textit{neurons}.
The response may be represented by various \textit{activation functions}, such as \ttt{sigmoid} or \ttt{tanh} functions.

\begin{figure}[tb]
\begin{center}
  \begin{minipage}[c]{0.5\textwidth}\begin{center}
    \includegraphics[trim=20mm 15mm 27mm 15mm,clip,width=0.75\textwidth]{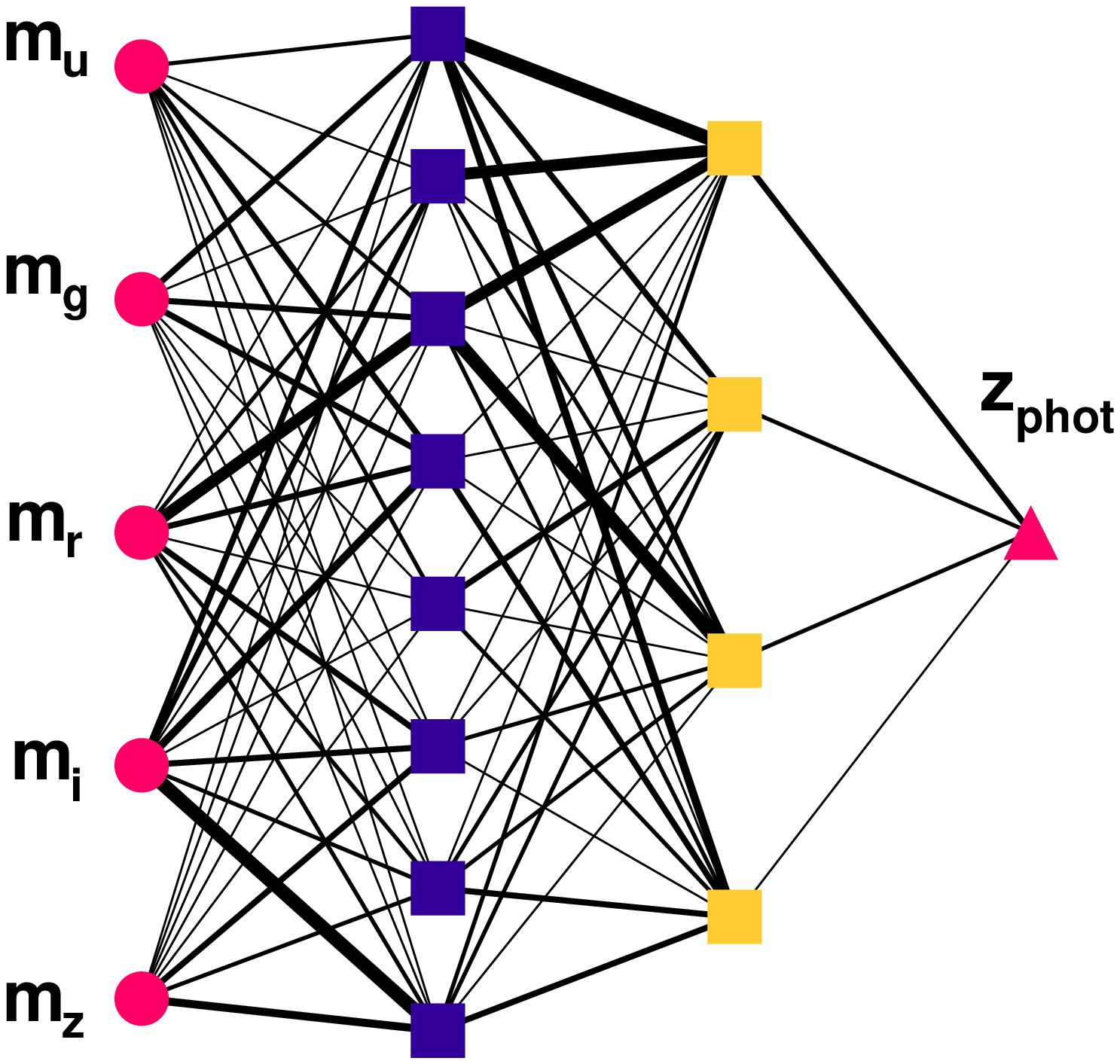}
  \end{center}\end{minipage}\hfill
  \caption{\label{FIG_annScetch}
    Schematic representation of an artificial neural network, with individual
    neurons marked by circles, squares and a triangle.
    The input variables to the ANN are five magnitudes,
    $\ttt{m}_{\ttt{u}}$, $\ttt{m}_{\ttt{g}}$, $\ttt{m}_{\ttt{r}}$, $\ttt{m}_{\ttt{i}}$ and $\ttt{m}_{\ttt{z}}$ (red circles).
    These are fed into the first hidden layer (blue squares), and
    further propagated into a second hidden layer (yellow squares). Finally, the sum of the
    second hidden layer is combined into the output of the ANN, the \photoz, \zPhot (red triangle).
    In each stage, the response of the various neurons is summed using relative weights, which are
    represented by the thickness of the interconnecting lines. The result of training an ANN
    is an optimized set of weights; for these, the response of the ANN recovers the desired mapping between
    the input variables and the target value or type, respectively for regression or classification.
  }
\end{center}
\end{figure}

In \annz, the \tmva method for ANNs called a \textit{multi-layer perceptron} is implemented. In this case,
ANN neurons are organized into at least three layers, the \textit{input layer}; a \textit{hidden layer};
and the \textit{output layer}, where more complicated structures may include multiple hidden layers.
A schematic illustration is shown in \autoref{FIG_annScetch}.

In the perceptron, the response of a neuron is fed into the next layer (up to the output), using a series of relative weights.
Learning occurs by changing the inter-neuron weights after each element
of the dataset is processed, using the so-called \textit{back propagation} algorithm.
This is carried out through a generalization of the least mean squares
algorithm, using the ANN \textit{error function}. The latter characterizes the amount of
error in the output compared to the predicted result in the validation dataset. 
In practice, the weights are varied using the gradient of the error function, though optionally,
the second derivatives of the error may also be used.

Using ANNs, it is important to avoid over-training. The latter occurs when an ANN becomes sensitive
to the fluctuations in a dataset, instead of to the coherent features of the observables which should be mapped to the output.
Over-training leads to a seeming increase in the performance, if measured on the training sample. Conversely,
it also results in an effective performance decrease, when measured from the independent validation sample.
Over-training may therefore be detected by comparing the value of the error estimator between the training and
the validation sample.
In addition to testing for over-training, \textit{convergence tests} may also be performed. The latter
refer to checking whether the estimator has ceased to improve over the course of several training cycles;
they are used in order to determine when to stop training

An additional feature available in \tmva is \textit{Bayesian regularization}. Regularization
adds a term to the error function of the ANN, which is equivalent to the negative value of
the log-likelihood of the training data, given the network model. Regularization reduces the
risk of over-training, by penalizing ANNs with over-complicated architectures (too many degrees of
freedom).

\begin{figure}[tb]
\begin{center}
  \begin{minipage}[c]{0.5\textwidth}\begin{center}
    \includegraphics[trim=0mm 18mm 0mm 5mm,clip,width=1.0\textwidth]{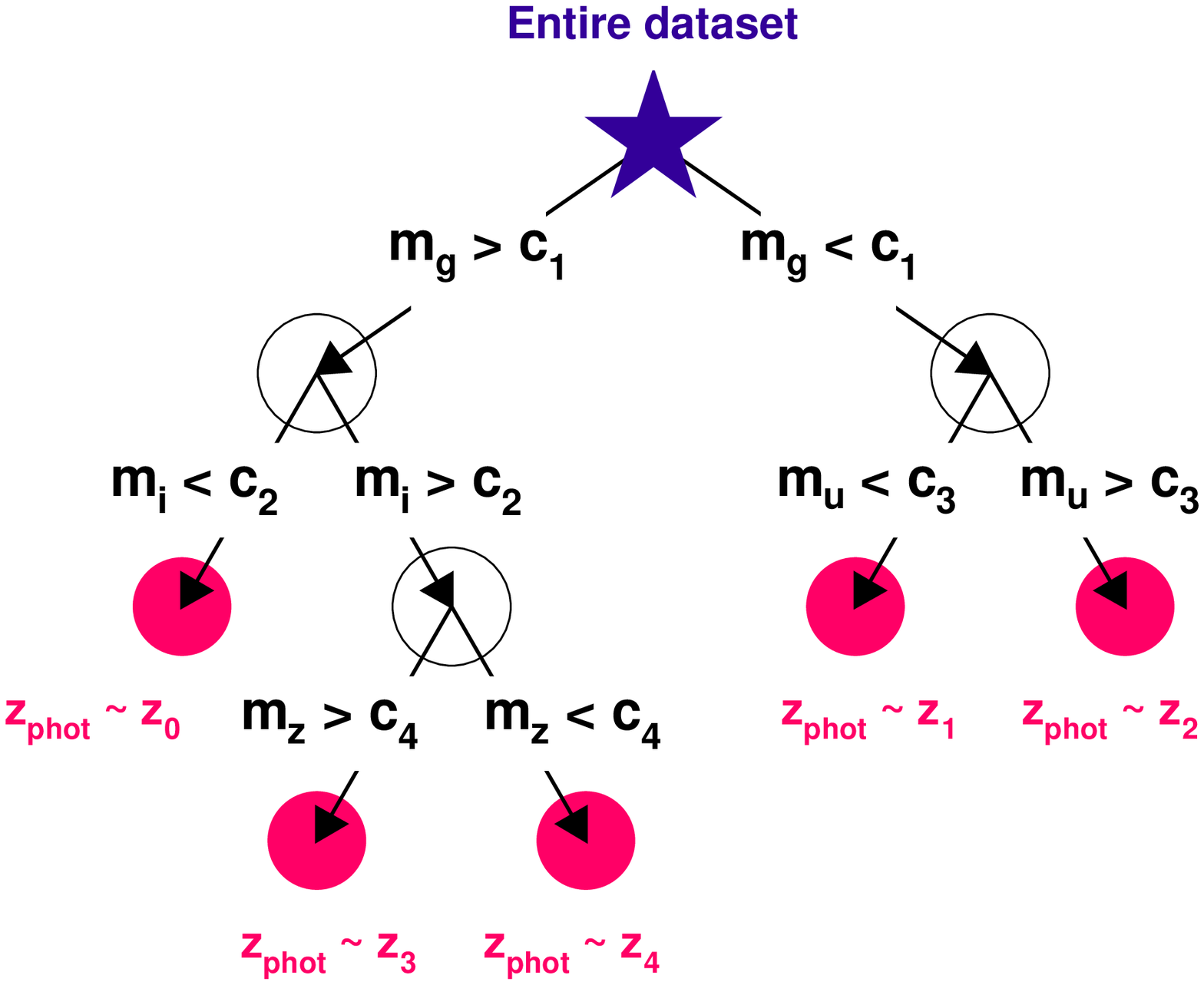}
  \end{center}\end{minipage}\hfill
  \caption{\label{FIG_bdtScetch}
    Schematic representation of a decision tree, with the initial root node marked by
    a star, internal nodes marked by empty circles, and output nodes (leafs) marked by full circles.
    A sequence of binary splits using magnitudes, $\ttt{m}_{\ttt{u}}$, $\ttt{m}_{\ttt{g}}$,
    $\ttt{m}_{\ttt{i}}$ and $\ttt{m}_{\ttt{z}}$, as input variables,
    is applied to each element of the training dataset.
    Each split uses the variable that, at that particular node, results in the best discrimination when being cut on.
    The leafs represent a division of the dataset into sub-samples in the target variable.
    In the case of regression, as in this example, these are associated with different values of
    the \photoz, \zPhot, denoted here by~$c_{1,2,3,4}$.
    For classification, each leaf represents a sub-set of signal- or of background-enriched examples.
  }
\end{center}
\end{figure}

\vspace{5pt}\noindent\headFont{-~Boosted decision trees (BDTs).}
A decision or regression tree\footnote{~We use here the terms decision- and regression-trees interchangeably.}
is a binary tree, in which decisions are taken on one single variable at
a time, until a stop criterion is fulfilled.
The decision tree splits the parameter space into a large number of hypercubes.
Each of these is attributed a constant target value for regression,
or identified as either ``signal-like'' or ``background-like'' for classification.
The various output nodes are referred to as \textit{leafs}.
The path down the tree to each leaf represents
an individual cut sequence that narrows down the value of the
regression target, or the identification as signal or as background.
A schematic representation of a decision tree is shown in \autoref{FIG_bdtScetch}.

The training, or \textit{growing}, of a decision tree is the process that defines the splitting criteria
for each node, the purpose of which is to achieve the best estimation of the regression target,
or the best separation between signal and background objects.
The training starts with the root node, which is split into two subsets of training objects.
In each subsequent step, further splitting occurs. 
At each node, the split is determined by finding the variable and corresponding cut value that provide the best 
discriminatory power.
Training stops when the minimum number of training examples
in a single leaf is reached, according to a predefined threshold value.
For regression, each leaf corresponds to the value of the regression target of the associated training examples.
For classification, a leaf is interpreted as signal or as
background, based on the type of the majority of corresponding examples.
Different splitting criteria can be selected by the user in \annz,
among other algorithm parameters.

Decision trees are sensitive to statistical fluctuations in the
training sample. This comes about,
as a small change in a single node may affect all subsequent nodes, and the entire
structure of the tree thereafter. It is therefore beneficial to use not a single
tree classifier, but a \textit{forest} of trees, by using a \textit{boosting} algorithm. The
process of boosting involves training multiple classifiers using the same data
sample, where the data are reweighted differently for each tree.
The combined estimator is then derived from the weighted majority vote of trees in the forest.
Alternatively, it is also possible to use \textit{bagging} instead of boosting. In the
bagging approach, a re-sampling technique is used; a classifier is repeatedly trained using
re-sampled training objects such that the combined classifier represents an average of the individual
classifiers.
Several boosting/bagging algorithms are implemented in \tmva, all available through \annz.

\vspace{5pt}\noindent\headFont{-~Other methods.}
%
The \tmva package includes several other machine learning methods which are not
discussed here, such as k-nearest neighbours,
support vector machines, multidimensional likelihood estimators
and function discriminant analysis. All of these are interchangeable in \annz;
the user may choose any type or combination of types of MLM, in order to derive
single-value solutions and PDFs.

\subsubsection{Method selection and parameter tuning}
%
Different MLMs have their own strengths and weaknesses. For instance,
the training of a BTD is generally much faster than \eg that of an ANN;
conversely, the evaluation time of ANNs is generally shorter than that for large random forests.
In order to select the best estimator for a given problem, it is recommended to derive solutions using multiple
methods, using various choices of algorithm parameters. This is done in an automated fashion in \annz.

\section{Example analysis}
%
The \annz package is provided with a small dataset, used as a toy analysis.
The data consist of observations of galaxies and stars, included in
the tenth data release (DR10) of the Sloan Digital Sky Survey (SDSS)~\citep{dr10.1307.7735},
including measurements taken with the Baryon Oscillation Spectroscopic Survey (BOSS)~\citep{boss.1208.0022}.

\begin{figure*}[tbp]
\begin{center}
  \begin{minipage}[c]{0.5\textwidth}
    \subfloat[]{\label{FIG_boss_inputDist0}\includegraphics[trim=-3mm 0mm 21mm 18mm,clip,width=.95\textwidth]{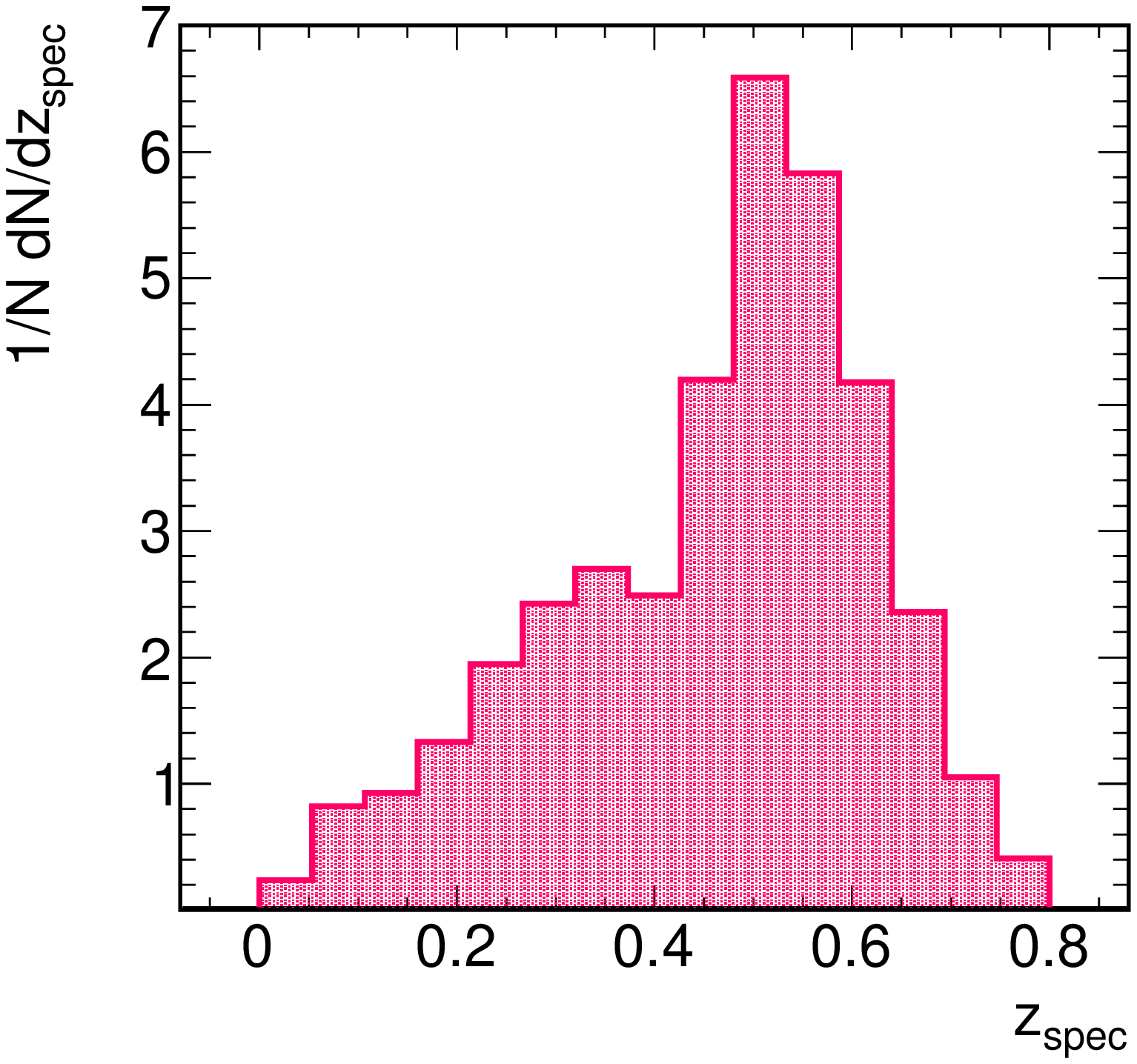}}
  \end{minipage}\hfill
  \begin{minipage}[c]{0.5\textwidth}
    \subfloat[]{\label{FIG_boss_inputDist1}\includegraphics[trim=0mm 0mm 19mm 18mm,clip,width=.95\textwidth]{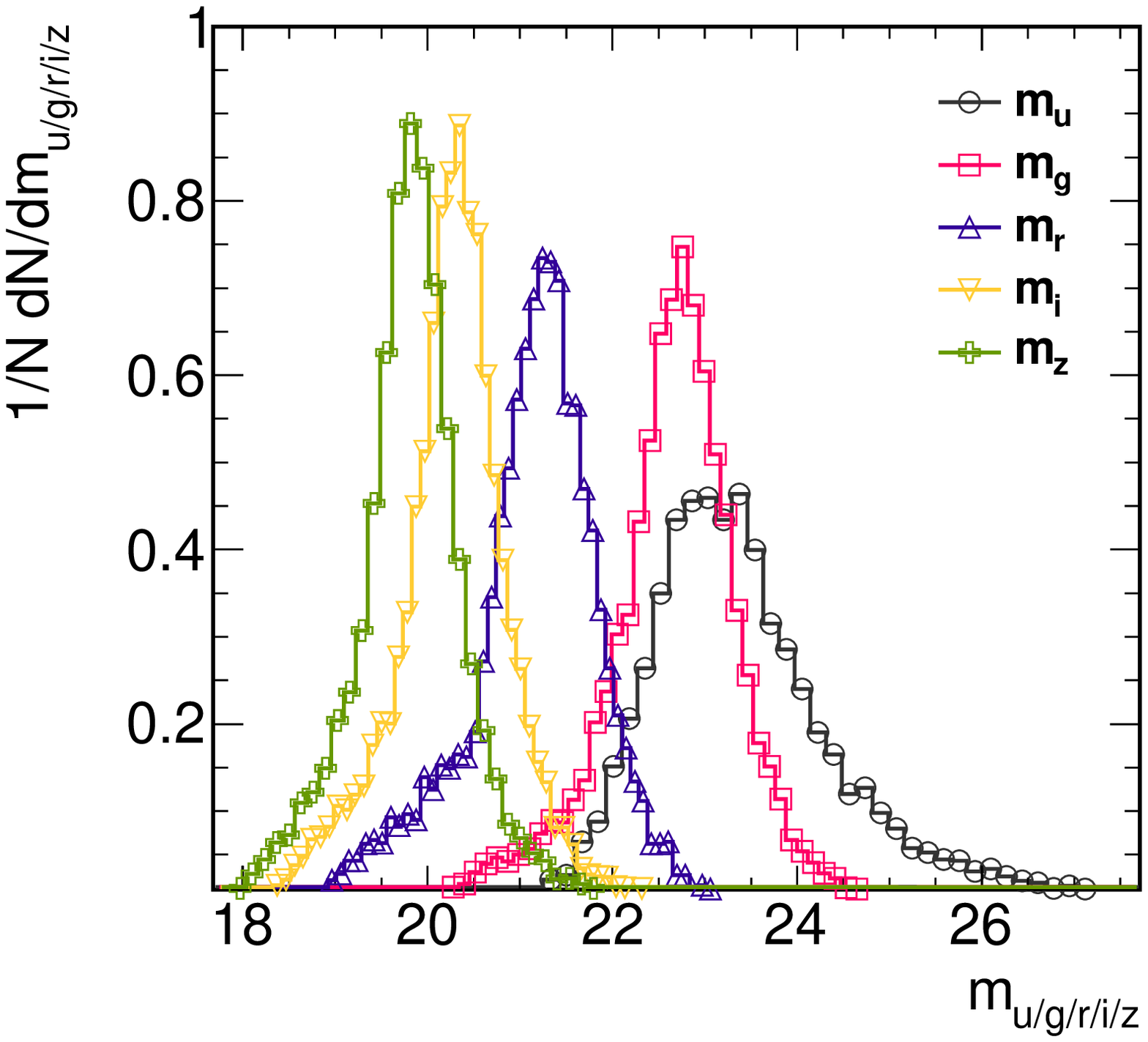}}
  \end{minipage}\hfill
  \begin{minipage}[c]{0.33\textwidth}
    \subfloat[]{\label{FIG_boss_inputDist2}\includegraphics[trim=5mm 0mm 30mm 15mm,clip,width=.95\textwidth]{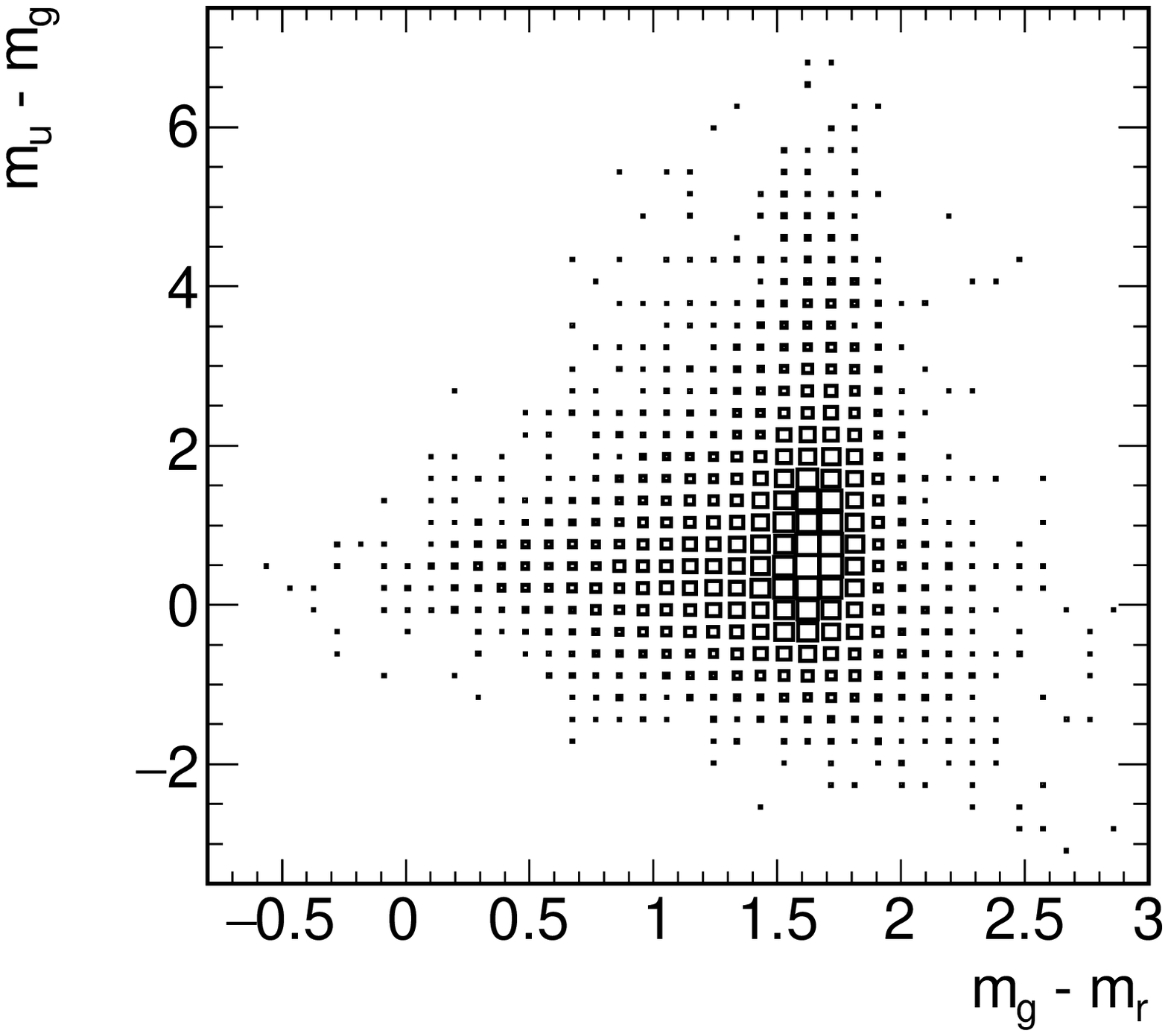}}
  \end{minipage}\hfill
  \begin{minipage}[c]{0.33\textwidth}
    \subfloat[]{\label{FIG_boss_inputDist3}\includegraphics[trim=5mm 0mm 30mm 15mm,clip,width=.95\textwidth]{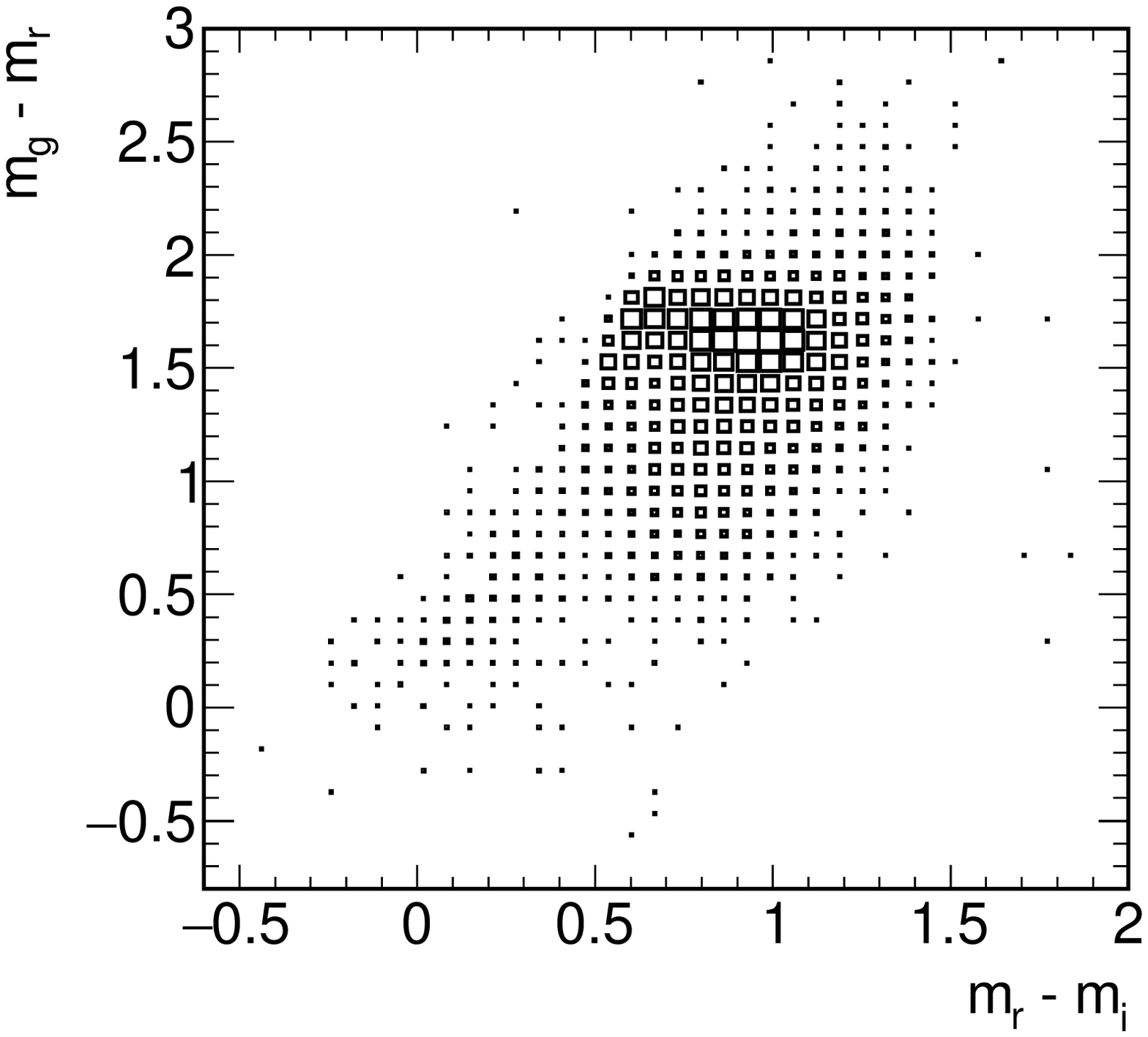}}
  \end{minipage}\hfill
  \begin{minipage}[c]{0.33\textwidth}
    \subfloat[]{\label{FIG_boss_inputDist4}\includegraphics[trim=5mm 0mm 30mm 15mm,clip,width=.95\textwidth]{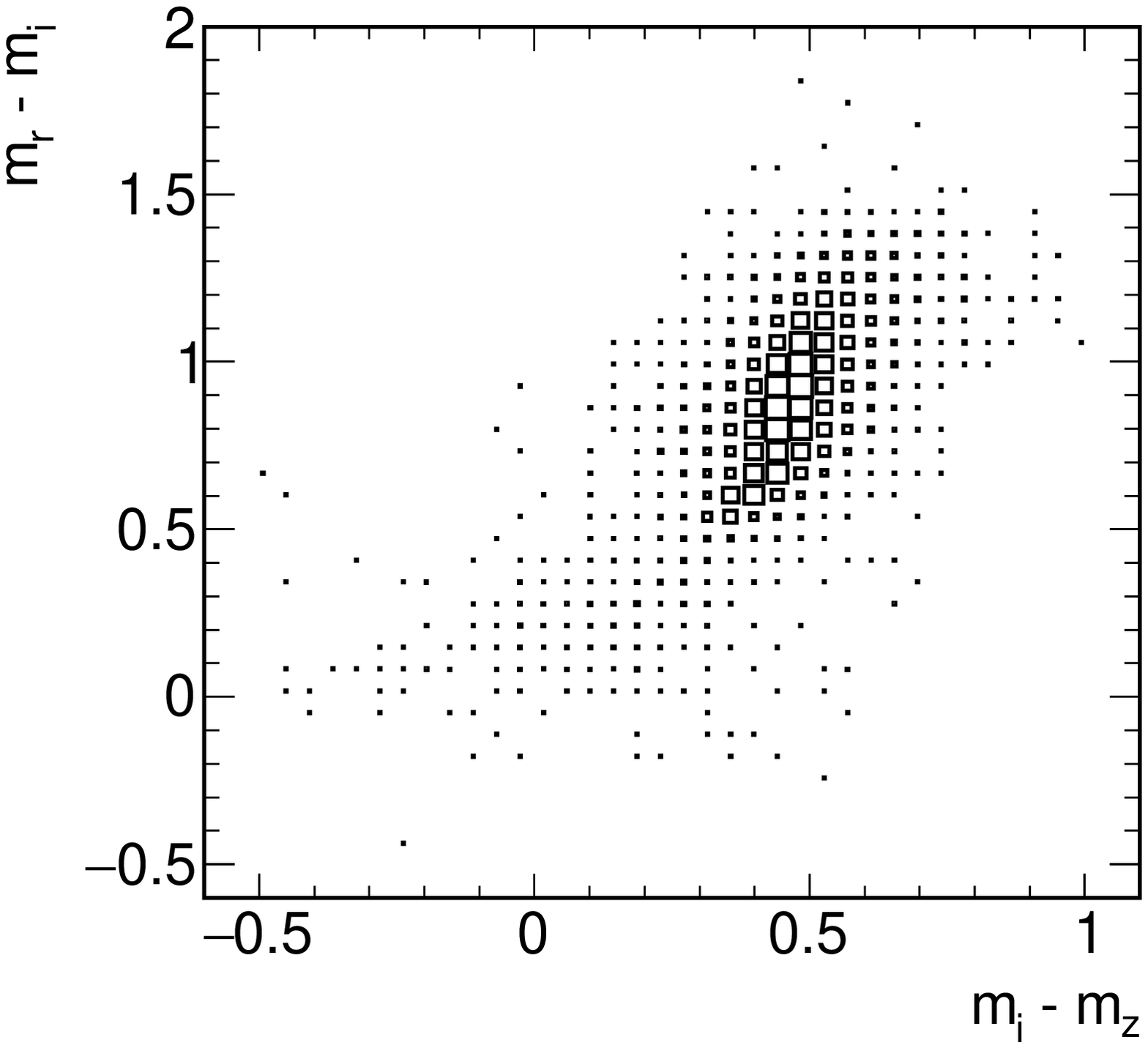}}
  \end{minipage}\hfill
  \caption{\label{FIG_boss_inputDist}
    Properties of galaxies in the dataset used for the toy \photoz analysis.
    \Subref{FIG_boss_inputDist0}~:~Differential distribution of the spectroscopic redshift, \zSpec.
    \Subref{FIG_boss_inputDist1}~:~Differential distributions of the magnitudes in five
    bands, $\ttt{m}_{\ttt{u}}$, $\ttt{m}_{\ttt{g}}$, $\ttt{m}_{\ttt{r}}$, $\ttt{m}_{\ttt{i}}$ and $\ttt{m}_{\ttt{z}}$, as indicated.
    \Subref{FIG_boss_inputDist2}-\Subref{FIG_boss_inputDist4}~:~Correlation between different colour combinations, as indicated,
    where the size of a box represents the relative number-density of entries within the respective histogram bin, compared to the entire distribution.
  }
\end{center}
\end{figure*}

The galaxy sample used for the \photoz analysis is derived from a publicly
available catalogue.\footnote{~See
\href{http://www.sdss3.org/dr10/spectro/spectro_access.php}{http://www.sdss3.org/dr10/spectro}~.}
The inputs for the \photoz inference are Pogson galaxy magnitudes in five
bands (\ttt{ugriz}). The magnitudes, \ttt{m}, were calculated from the provided flux measurements, \ttt{f},
using the relation, ${\ttt{m} = 22.5 - 2.5 \log_{10}(\ttt{f}) \;.}$
The general properties of the dataset, comprised of roughly ${180\mrm{k}}$~objects, are shown in \autoref{FIG_boss_inputDist}.
%

\section{Definition of metrics and notation}\label{SECdeefinitionOfMetrics}
%
In order to quantify the performance of the different configurations of \annz, several metrics are used. 
The metrics serve both as part of the dynamic optimization procedure of \annz, and as a means
of assessing the quality of the results.
All calculations take into account per-object weights. Weights may be defined by the user, or derived
on the fly based on the type of analysis. For instance, the user may choose to down-weight certain galaxies based
on an associated degree of confidence. Such a sub-sample would then have lower relative significance
during training and optimization. Weights are also used in order to account for unrepresentative training
samples, as described in \autoref{SECrepresentativenessAndReliability}.

The following metrics are used.
The \textit{photometric bias} of a single galaxy is defined as
${\delta_{\mrm{gal}} = z_{\mathrm{phot}} - z_{\mathrm{spec}}}$,
where $z_{\mathrm{phot}}$ and
$z_{\mathrm{spec}}$ are respectively the photometric and spectroscopic redshifts of the galaxy.
The \textit{photometric scatter} represents the standard deviation of $\delta_{\mrm{gal}}$ for a collection of
galaxies. Similarly,
$\sigma_{68}$ denotes the half-width of the area enclosing the peak \sixEightP of the distribution
of $\delta_{\mrm{gal}}$.
Another useful qualifier is the \textit{outlier fraction} of the bias distribution, $f(\alpha\sigma)$, defined as
the percentage of objects which have a bias larger than some factor, $\alpha$, of either $\sigma$ or $\sigma_{68}$.
In addition, we also define the \textit{combined outlier fraction} for~2 and~$3\sigma_{68}$,
${f(2,3\sigma_{68}) = \frac{1}{2} \left(\; f(2\sigma_{68}) + f(3\sigma_{68}) \;\right)}$~.

The various metrics are calculated for galaxies in bins of either \zPhot or \zSpec, and are
denoted in the following by a subscript, $b$, as $\delta_{b}$, $\sigma_{b}$, $\sigma_{68,b}$ and $f_{b}(\alpha\sigma)$.
The average values of the metrics over all redshift bins are denoted
by $<\delta>$, $<\sigma>$, $<\sigma_{68}>$ and $<f(\alpha\sigma)>$,
and serve as single-value qualifiers of the entire sample of galaxies.

The purpose of the bias, scatter and outlier fractions is to qualify the galaxy-by-galaxy \photoz
estimation. Additionally, the overall fit of the photometric redshift distribution, $N(\zPhot)$,
to the true redshift distribution, $N(\zSpec)$, is assessed using two
metrics. The first is denoted by $N_{\mrm{pois}}$, and stands for the sum of the 
bin-wise difference between the two distributions, normalized by the Poissonian fluctuations.
The second measure is the value of the \textit{Kolmogorov-Smirnov} (KS) test of $N(\zPhot)$ and $N(\zSpec)$,
which stands for the maximal distance between the cumulative distribution functions of the two distributions.
The $\mrm{KS}$-test has the advantage that, unlike $N_{\mrm{pois}}$,
it does not depend on the choice of binning of the redshift distributions.
The absolute value of the $N_{\mrm{pois}}$ and $\mrm{KS}$-test statistics is not
necessarily significant. Rather, these serve to compare the compatibility of
the \zPhot and \zSpec distributions, between different \photoz estimators.

\section{The \annz algorithm}\label{SECdescriptionOfTheAlgorithm}
%
\subsection{\Photoz PDF derivation}
%
The primary configurations of \annz are referred to as \textit{single regression} and \textit{randomized regression}.
These are respectively used to derive single-value solutions and PDFs.
The PDFs provided by \annz are intended to provide a description of
our knowledge of the \photoz solution.
Assuming one could reconstruct a perfect photometric redshift, the corresponding PDF would
be given by a delta function. However, the redshift inference has intrinsic uncertainties.
A \photoz PDF can thus be thought of as a way to parametrize the uncertainty on the solution.

The main contributing factors to the uncertainty on \photozs are the following:

\begin{enumerate}{}
\item
\textbf{$(\mathcal{U}_{1})$~Uncertainty on inputs to training:} magnitudes are not
sufficient to derive the redshift, as they only provide
a rough sampling of the underlying SED.  Furthermore, one also needs
to consider the uncertainties on the values of the magnitudes.
The latter are usually derived from
the Poissonian noise on the corresponding photon-count, and so are
under-estimated. These uncertainties are therefore
difficult to take into account in the \photoz derivation in a direct way.
\item
\textbf{$(\mathcal{U}_{2})$~Uncertainty on MLMs:} there is an inherent uncertainty on the solution of a given MLM.
For example, different initial
random seeds for training, or the choice of different MLM algorithms, may result in variation in
the performance.
\item
\textbf{$(\mathcal{U}_{3})$~Unrepresentative training datasets:} the training data may
not be representative of the evaluated photometric sample. In this case, 
the results are influenced by the composition of the training dataset (the relative proportion
of training galaxies with different combinations of magnitudes).
\item
\textbf{$(\mathcal{U}_{4})$~Incomplete training datasets:} the training data may
not be complete. This may occur if some regions of magnitude-space,
which exist in the evaluated sample, have no corresponding galaxies for training. The
\photoz predictions for such evaluated galaxies are unreliable.
\end{enumerate}

Of these sources of uncertainty, the first three may be incorporated into a meaningful PDF.
The dominant effect of the latter
is the degenerate mapping between magnitudes and redshift~$(\mathcal{U}_{1})$.
As an example, one may consider the small gap between the response curves of
the SDSS~$\ttt{g}$- and~$\ttt{r}$-band filters. The latter results in an ambiguity in
the location of the ${4000~\mrm{\AA}}$ Balmer break between the two bands, for
galaxies with ${z\sim0.35}$~\citep{schmidt2007galaxy}. The degeneracy manifests itself as large \photoz
uncertainties for this redshift region, as \eg evident from \autoref{FIG_boss_annz1_annz2_cor0} below.

Glossing for the moment over the the technical details,
the procedure for deriving our PDF is as follows.
We start by producing a single-value \photoz solution.
We then combine this solution with the corresponding \photoz uncertainty due
to the training inputs~$(\mathcal{U}_{1})$, 
which is derived as explained in \autoref{SECsingleRegression}.
The procedure is repeated for an ensemble of MLM estimators. The MLMs differ from each other
in the choice of algorithm and of algorithm settings, \eg numbers of neurons in an ANN, number
of trees in a BDT and so forth~$(\mathcal{U}_{2})$.
The various estimators and their corresponding uncertainties are then combined
into a PDF, as detailed in \autoref{SECrandomRegression}.

In general, the variance between different estimators is sub-dominant compared to the
\photoz uncertainty on a single MLM. However, the combination of different estimators allows
for the reconstruction of multi-peak PDFs, exposing degeneracies.
This comes about, as each MLM is sensitive to different statistical fluctuations.
Subsequently, each MLM has a slightly different response in cases where the \photoz/redshift relation is
ambiguous.
Using multiple MLMs also has the advantage of exposing configurations which perform
badly due to a poor choice of algorithm parameters, or to a statistical fluctuation in
the training. Conversely, consider an example where \eg a pair of
ANNs with different numbers of neurons exhibit slightly different performance.
Combining several solutions takes away some of the arbitrariness of selecting one
specific model.

The uncertainties on the make-up of the training dataset~$(\mathcal{U}_{3},\;\mathcal{U}_{4})$
can only partially be addressed. To deal with unrepresentative training samples, we employ training weights. The
latter are used to match the distribution of the inputs (\eg magnitudes)
from the training sample, to those from the evaluated
data~\citep{Lima:2008kn}. The calculation of the weights is performed as part of the internal pipeline of the code.
The issue of incomplete training samples can not be taken into account without the use of
additional data (such as those derived from simulations or from template libraries).
\annz therefore provides a quality
flag, which indicates when unrepresented data are being evaluated. A short discussion
is given in \autoref{SECrepresentativenessAndReliability}.

An alternative type of PDF is also generated by \annz, using the
\textit{binned classification} configuration.
This approach has been used in the past, following the
methodology of~\cite{Gerdes:2009tw}.
In binned classification, we build up a PDF by estimating the local \photoz probability in
narrow redshift regions, implementing classification MLMs instead of regression.
We have found that this method tends to
under-perform compared to randomized regression. Binned classification is therefore
not discussed here further, though an example analysis is provided with the software package.

In the next sections, we describe in detail the \annz algorithm.
All figures in the following are based on testing data (galaxies which were not used
as part of the training/validation phase).

\subsection{Single regression and uncertainty estimation}\label{SECsingleRegression}
%
In the simplest configuration of \annz, a single regression is performed.
This is similar to the nominal product of the original version of the code, \annzOne.

\begin{figure*}[tb]
\begin{center}
  \begin{minipage}[c]{0.5\textwidth}
    \subfloat[]{\label{FIG_boss_annz1_annz2_cor0}\includegraphics[trim=5mm 0mm 14mm 18mm,clip,width=1.0\textwidth]{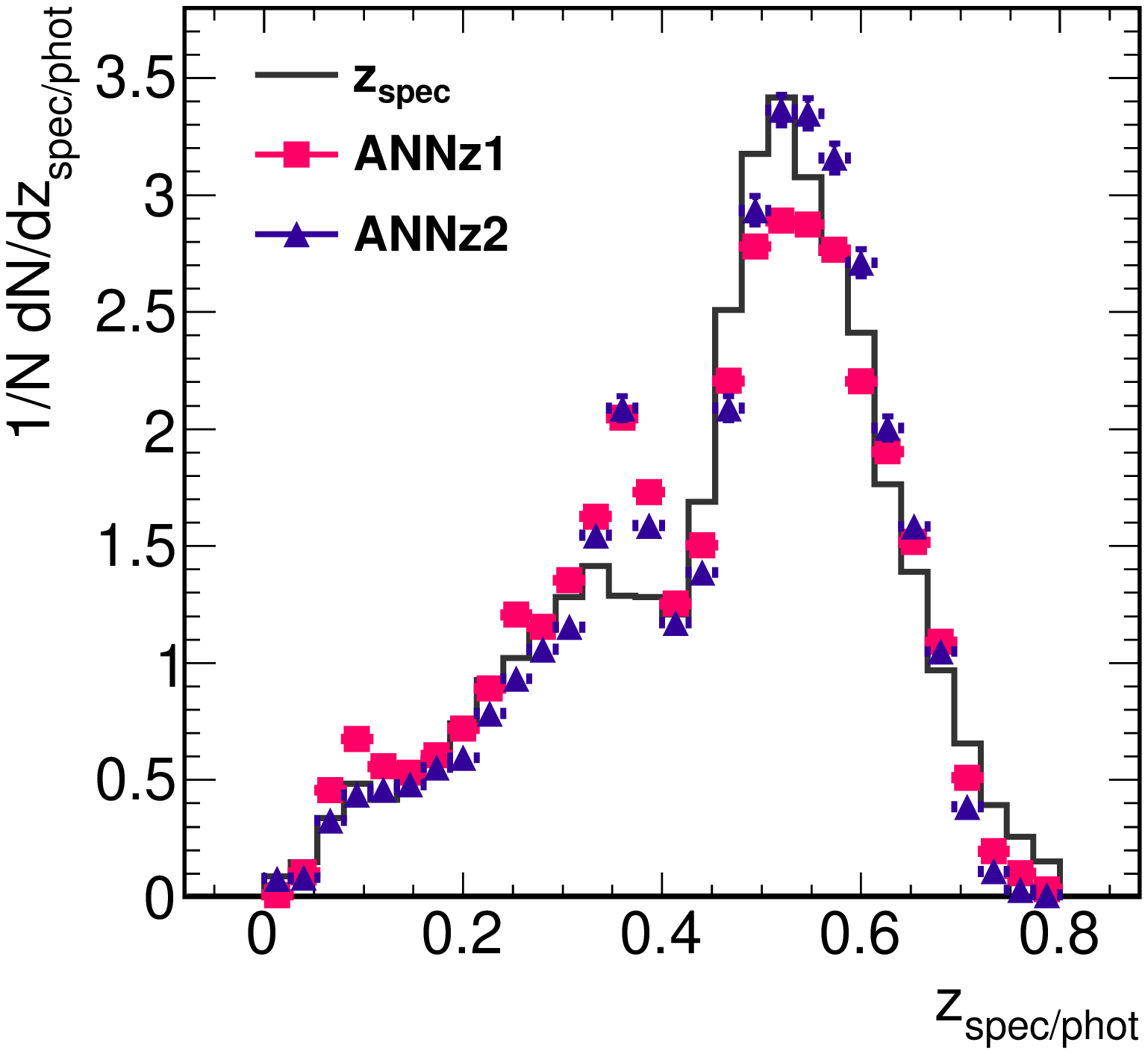}}
  \end{minipage}\hfill
  \begin{minipage}[c]{0.5\textwidth}
    \subfloat[]{\label{FIG_boss_annz1_annz2_cor1}\includegraphics[trim=5mm 0mm 14mm 18mm,clip,width=1.0\textwidth]{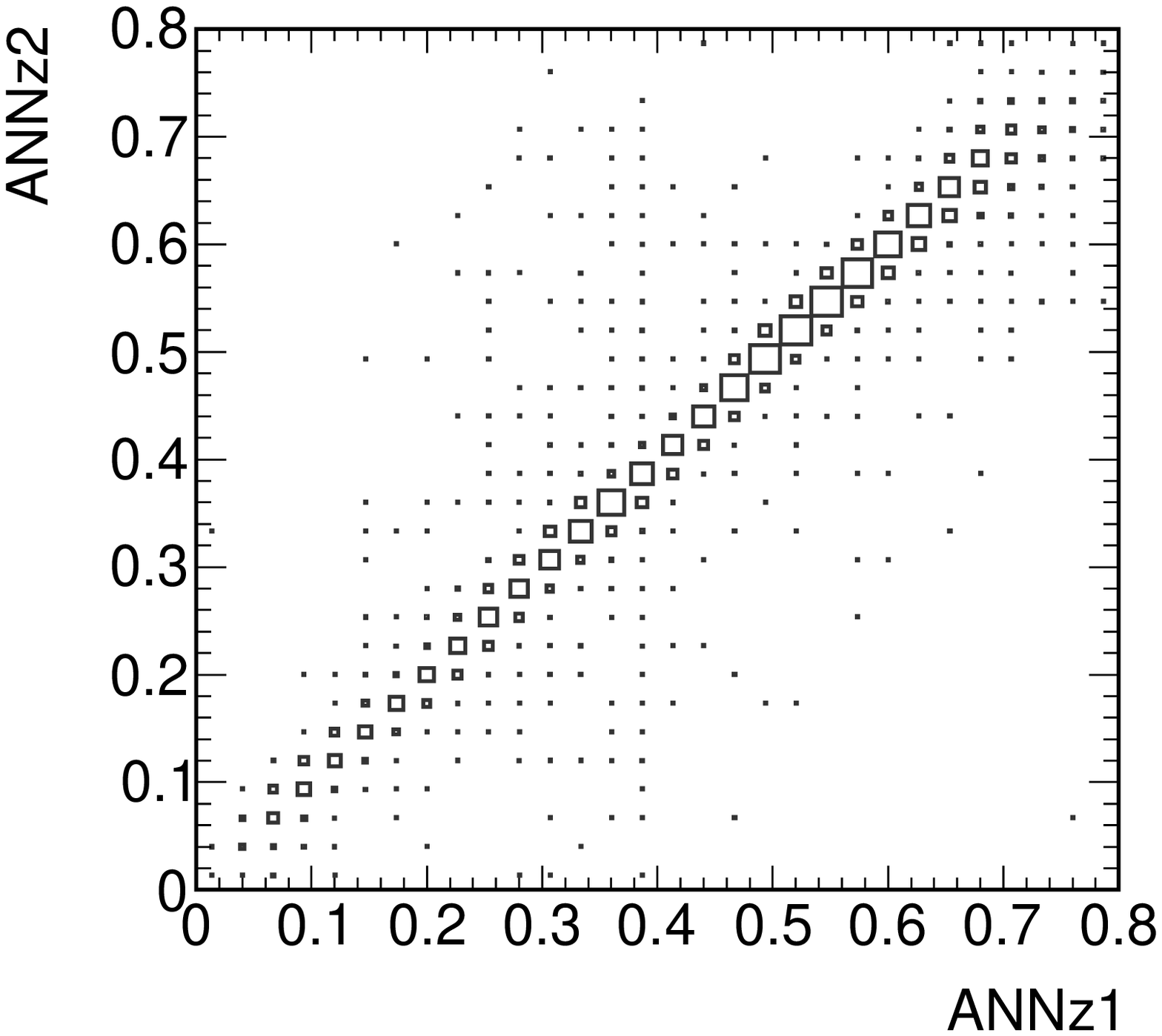}}
  \end{minipage}\hfill
  \caption{\label{FIG_boss_annz1_annz2_cor}
    Comparison between the \photoz solutions of \annz and the original version of the
    code, derived using a single ANN, as described in the text.
    \Subref{FIG_boss_annz1_annz2_cor0}~:~Differential distributions of the spectroscopic
    and the photometric redshift, respectively \zSpec and \zPhot, of \annzOne and \annz, as indicated.
    \Subref{FIG_boss_annz1_annz2_cor1}~:~Correlation between the \photoz solutions of \annzOne and \annz.
    Around~$z=0.35$, We observe a mismatch between the two estimators and \zSpec, as well as an
    increase in the scatter between the two. This indicates that the uncertainty on the
    \photozs in this region is large. The latter is difficult to reconcile using single-value estimators,
    but is alleviated using a PDF, as discussed in the text (also see \autoref{boss_NzNominal}).
  }
\end{center}
\end{figure*}

We compare the output of \annz with that of \annzOne in \autoref{FIG_boss_annz1_annz2_cor}.
In both cases, a single ANN with architecture $\{N,N+1,N+9,N+4,1\}$ was used;
this corresponds to $N=5$ input parameters (five magnitudes) in the first layer,
three hidden layers with various numbers of neurons, and one output neuron
in the final layer.\footnote{~This network architecture was
found to produce optimal performance for our particular dataset, and
is denoted below as \zBest. However, for a different analysis, another architecture might be preferred.}
A sample of ${30\mrm{k}}$ objects was used for the training. Comparable results
were also achieved, using as many as ${200\mrm{k}}$, and as few as ${5\mrm{k}}$ objects.

The redshift distributions derived
by the two versions of the code are similar, with somewhat better performance of \annz over the original version.
One may notice the large uncertainty on the \photozs around~$z=0.35$ for both estimators, as mentioned above.
Such discrepancies between the derived \photozs and the true redshift are difficult to reconcile using a
single-value MLM. However, a PDF solution helps to alleviate the problem (see \autoref{boss_NzNominal} below).
In order to understand how to derive a PDF, we must first qualify the performance of a single MLM.

\begin{figure*}[tbp]
\begin{center}
  \begin{minipage}[c]{0.5\textwidth}
    \subfloat[]{\label{FIG_boss_singleANNmetrics0}\includegraphics[trim=5mm 0mm 14mm 18mm,clip,width=1.0\textwidth]{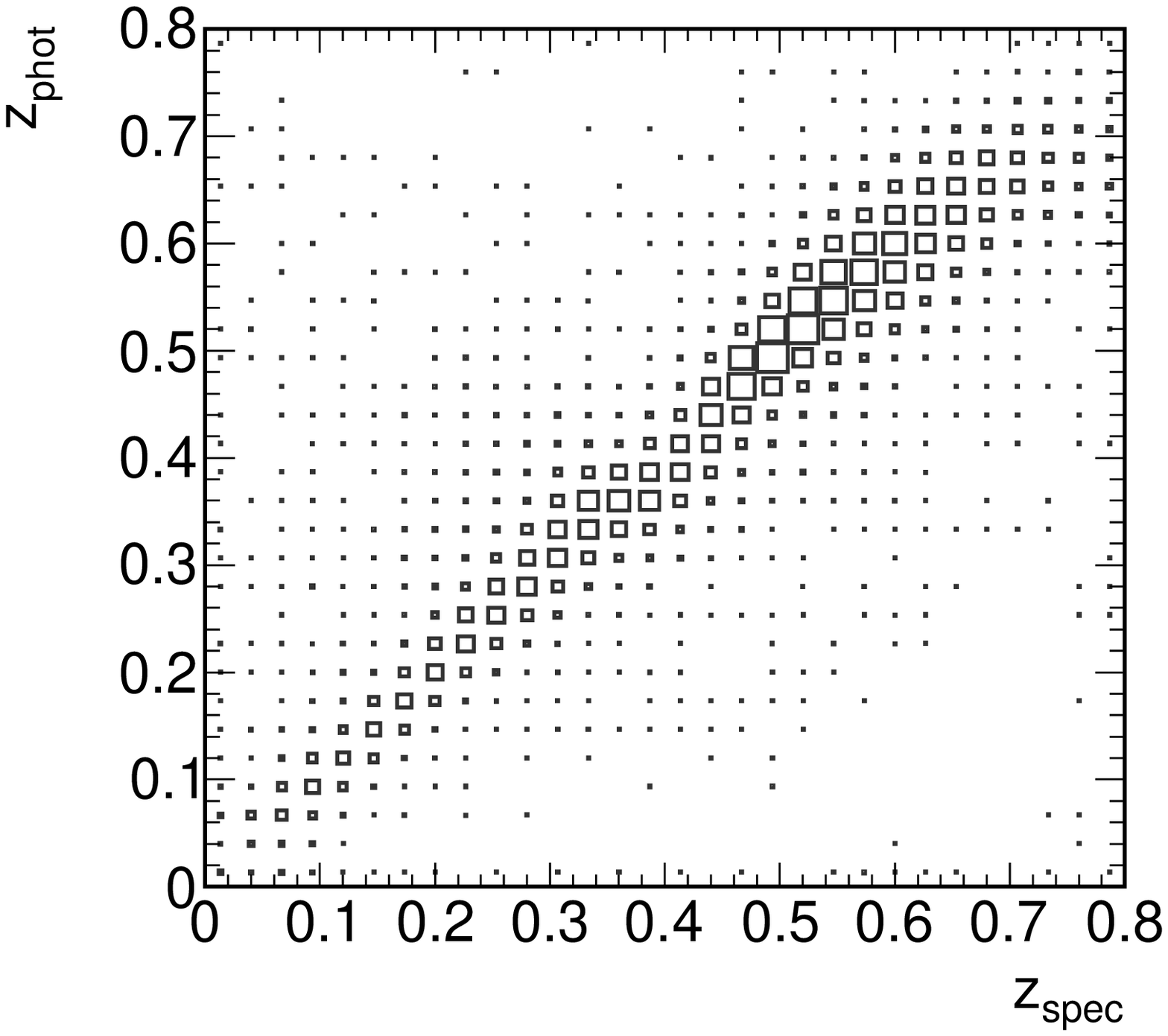}}
  \end{minipage}\hfill
  \begin{minipage}[c]{0.5\textwidth}
    \subfloat[]{\label{FIG_boss_singleANNmetrics1}\includegraphics[trim=5mm 0mm 14mm 18mm,clip,width=1.0\textwidth]{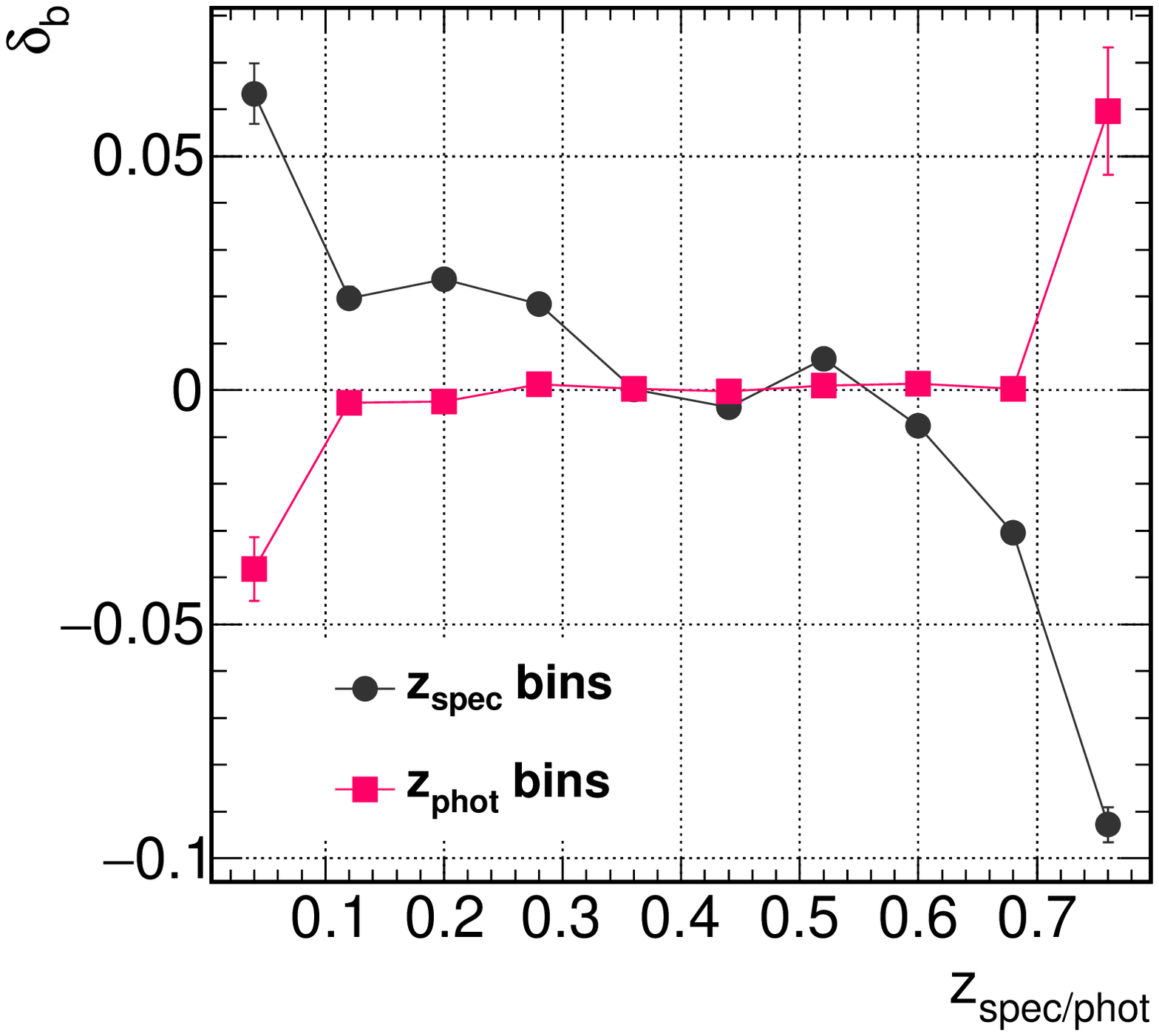}}
  \end{minipage}\hfill
  \begin{minipage}[c]{0.5\textwidth}
    \subfloat[]{\label{FIG_boss_singleANNmetrics2}\includegraphics[trim=5mm 0mm 14mm 18mm,clip,width=1.0\textwidth]{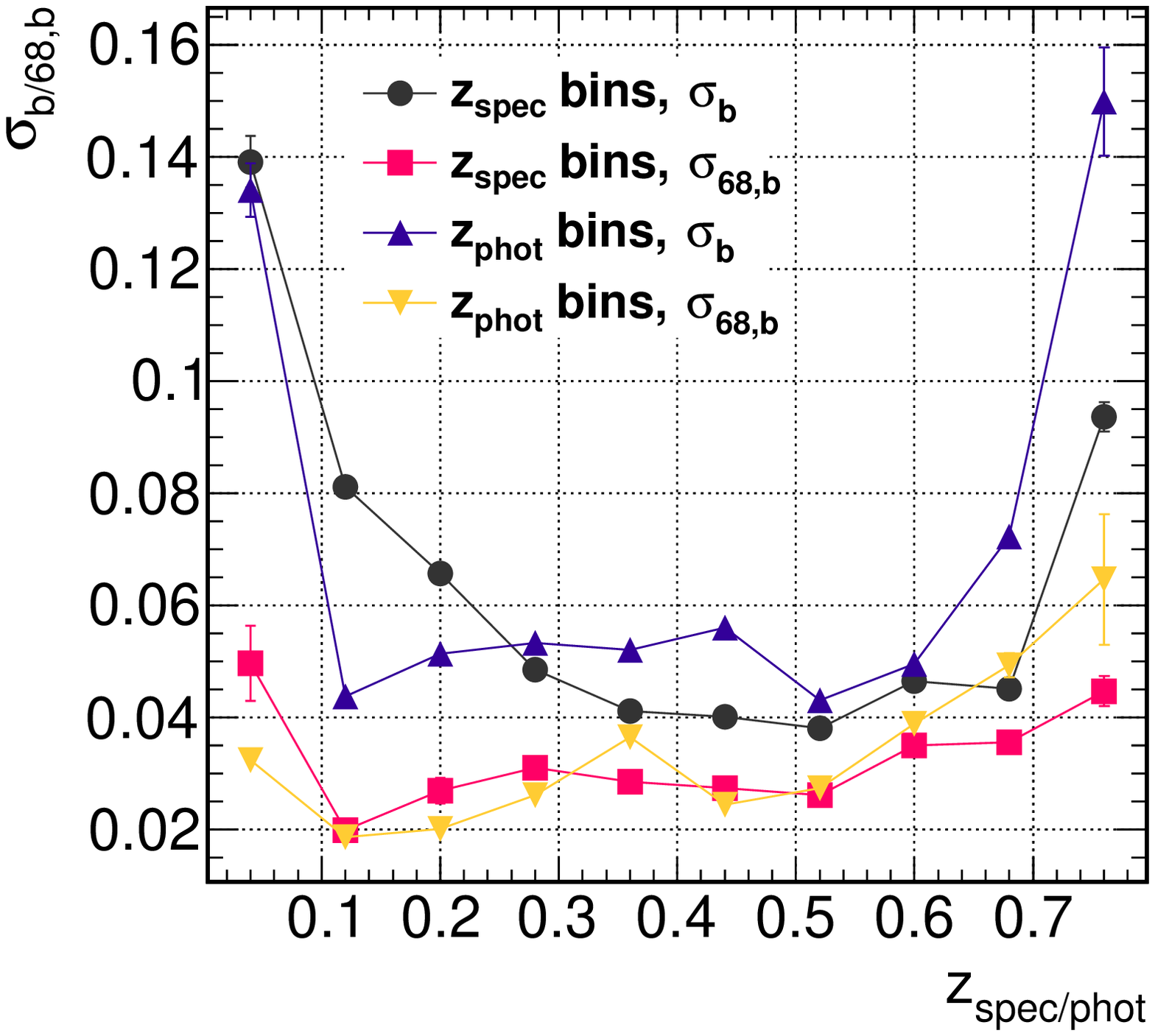}}
  \end{minipage}\hfill
  \begin{minipage}[c]{0.5\textwidth}
    \subfloat[]{\label{FIG_boss_singleANNmetrics3}\includegraphics[trim=5mm 0mm 14mm 18mm,clip,width=1.0\textwidth]{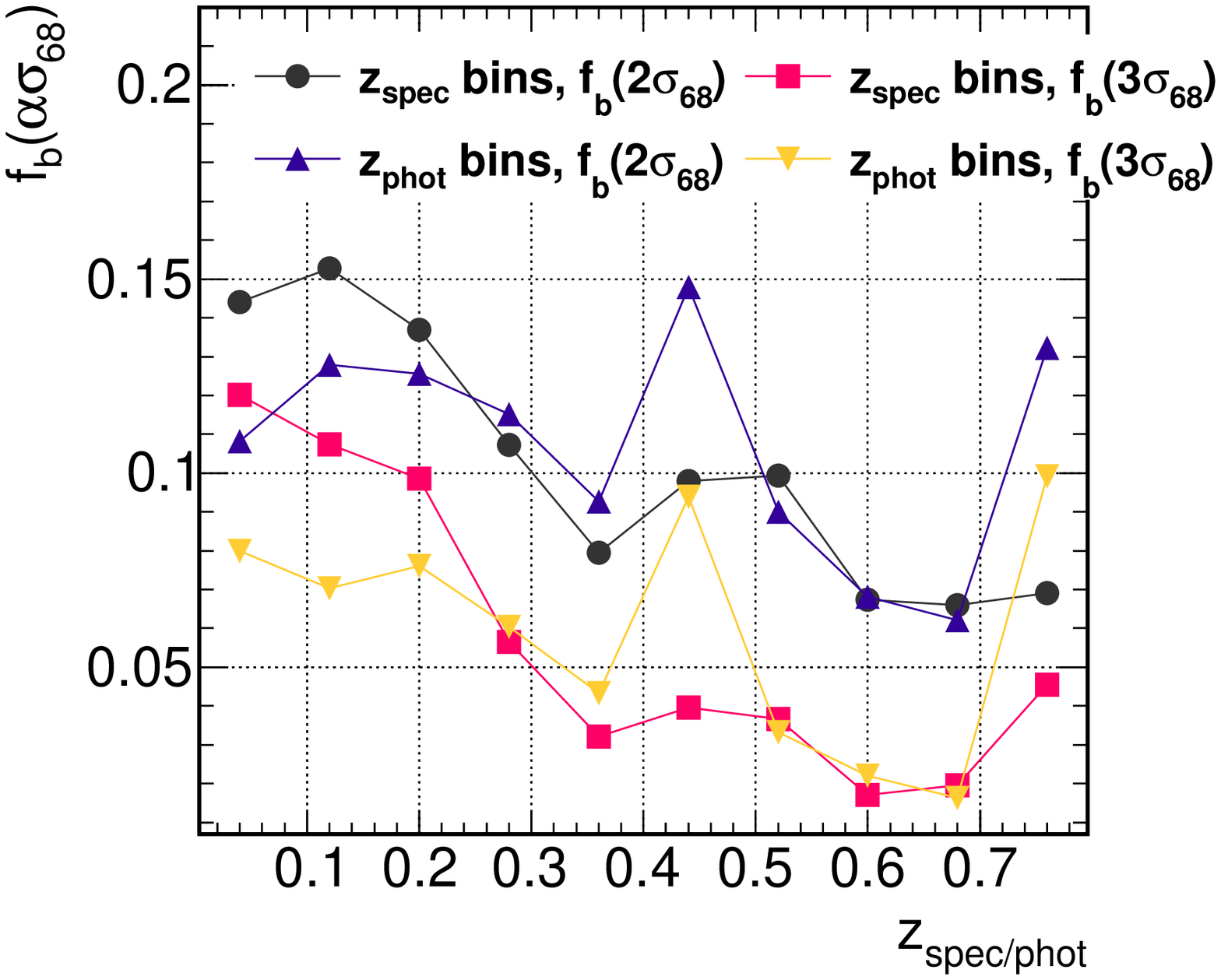}}
  \end{minipage}\hfill
  \caption{\label{FIG_boss_singleANNmetrics}
    Properties of the \photoz solution of \annz, derived using a single ANN, as described in the text.
    \Subref{FIG_boss_singleANNmetrics0}~:~Correlation between the spectroscopic and the photometric
    redshift, respectively \zSpec and \zPhot.
    \Subref{FIG_boss_singleANNmetrics1}~:~The \photoz bias, $\delta_{b}$, calculated in bins of either
    \zSpec or \zPhot, as indicated.
    \Subref{FIG_boss_singleANNmetrics2}~:~The \photoz scatter, calculated as either the standard deviation
    or as the \sixEightP of the distribution of the bias,
    respectively $\sigma_{b}$ and $\sigma_{68,b}$, calculated in bins of either \zSpec or \zPhot, as indicated.
    \Subref{FIG_boss_singleANNmetrics3}~:~The \photoz outlier fraction, ${f_{\mrm{b}}(\alpha\sigma_{68})}$,
    using~${\alpha=2\; \mrm{or} \;3}$, calculated in bins of either \zSpec or \zPhot, as indicated.
    The lines in~\Subref{FIG_boss_singleANNmetrics1}~-~\Subref{FIG_boss_singleANNmetrics3} are meant to guide the eye.
  }
\end{center}
\end{figure*}

The relation between the spectroscopic redshift and the \photoz estimator
of \annz is shown in \autoref{FIG_boss_singleANNmetrics0}. 
We observe a strong correlation between \zSpec and \zPhot.
\Autorefs{FIG_boss_singleANNmetrics1}~-~\subref{FIG_boss_singleANNmetrics3},
show the \photoz bias, scatter and outlier fractions as a function of the true and of the derived redshift values of \annz.
All metrics exhibit worse performance at the edges of the redshift range, due in part to the relatively small number of
respective training objects.

An additional important quantity which characterizes the performance of the
code is the associated \photoz uncertainty. 
For \annzOne, uncertainties were derived using a chain rule,
propagating the uncertainties on the algorithm-inputs, to an uncertainty on the value of the final \photoz.
The disadvantage of such a scheme is that the uncertainty on photometric inputs,
such as magnitudes, is not always precise in itself. This is due to the fact that in most cases,
the available uncertainty estimation only represents the Poissonian noise on the corresponding photon-count.
It therefore does not take into account other systematic uncertainties or correlations between observables.

In order to compute the uncertainty associated with our \photoz estimator, denoted hereafter
as $\sigma_{\mathrm{gal}}$, a data-driven method is employed.
This is done by assuming that objects with similar combinations of photometric properties should also
have similar \photoz uncertainties.
We derive the uncertainty using the \textit{K-nearest neighbours} (KNN) method.
We would emphasise that the latter should not be confused
with K-nearest neighbours machine learning.
For the calculation of $\sigma_{\mathrm{gal}}$, no additional training of an
MLM is required. Rather, a simple search in parameter-space is performed.

For example, let us assume that magnitudes are used as inputs for training.
In this case, the \textit{distance} in parameter-space between a pair of galaxies, $x$ and $y$, can be defined as 
\begin{equation}
  R_{\mrm{NN}}(x,y) = \sqrt{ \sum\limits_{j} \left( m_{j}^{x}-m_{j}^{y} \right)^{2} } \;,
\label{knnDistanceEq}
\end{equation}
where the ${m_{j}^{x,y}}$ symbols stand for the five magnitudes,
$\ttt{m}_{\ttt{u}}$, $\ttt{m}_{\ttt{g}}$, $\ttt{m}_{\ttt{r}}$, $\ttt{m}_{\ttt{i}}$ and $\ttt{m}_{\ttt{z}}$,
for the two galaxies.
The first step in the calculation is to find the $n_{\mrm{NN}}$ nearest neighbours
to our \textit{target object}, defined as those with the smallest value of ${R_{\mrm{NN}}}$
from the entire training sample.
For each of the neighbours, we calculate the 
\photoz bias. For neighbour~$i$, the latter is defined as
${\delta_{\mrm{NN}}^{i} = z^{i}_{\mathrm{phot}} - z^{i}_{\mathrm{spec}}}$,
where $z^{i}_{\mathrm{phot}}$ is the estimated \photoz of the objects,
and $z^{i}_{\mathrm{spec}}$ is the respective spectroscopic redshift.
The \sixEightP width of the distribution
of $\delta_{\mrm{NN}}^{i}$ values is then taken as the uncertainty on the \photoz
of the target object, $\sigma_{\mathrm{gal}}$.\footnote{~In practice, we calculate the \photoz uncertainty
separately for shifts to lower or to higher values of redshift.
However, for the sake of brevity,
we refer to $\sigma_{\mathrm{gal}}$ as symmetric in the following.}

This technique has been shown to produce realistic photometric uncertainties, as \eg
in~\citet{Oyaizu:2007jw}, so long as the training dataset is representative of
the evaluated photometric sample. Additionally, the authors there discussed the optimal value for $n_{\mrm{NN}}$.
It was explained that on the one hand, $n_{\mrm{NN}}$
should be large enough that the uncertainty estimation is not
limited by shot noise; on the other hand, $n_{\mrm{NN}}$
should not be set too high, so that the estimate remains
relatively local in the input parameter space.
For the current study, a nominal value, ${n_{\mrm{NN}}=100}$, was selected.

We would like to assert that the uncertainty estimator represents the correct underlying \photoz scatter
in our analysis.
For this purpose, we define the metric
\begin{equation}
  \rho_{\mrm{NN}} = \frac{\delta_{\mrm{gal}}}{\sigma_{\mathrm{gal}}} \;,
\label{knnRelErrorEq}
\end{equation}
the ratio between the \photoz bias and the associated uncertainty.
The distribution of the values of $\rho_{\mrm{NN}}$ for the entire sample is expected to be centred close
to zero, and to have a width close to unity.

\begin{figure}[tb]
\begin{center}
  \includegraphics[trim=5mm 5mm 14mm 18mm,clip,width=.475\textwidth]{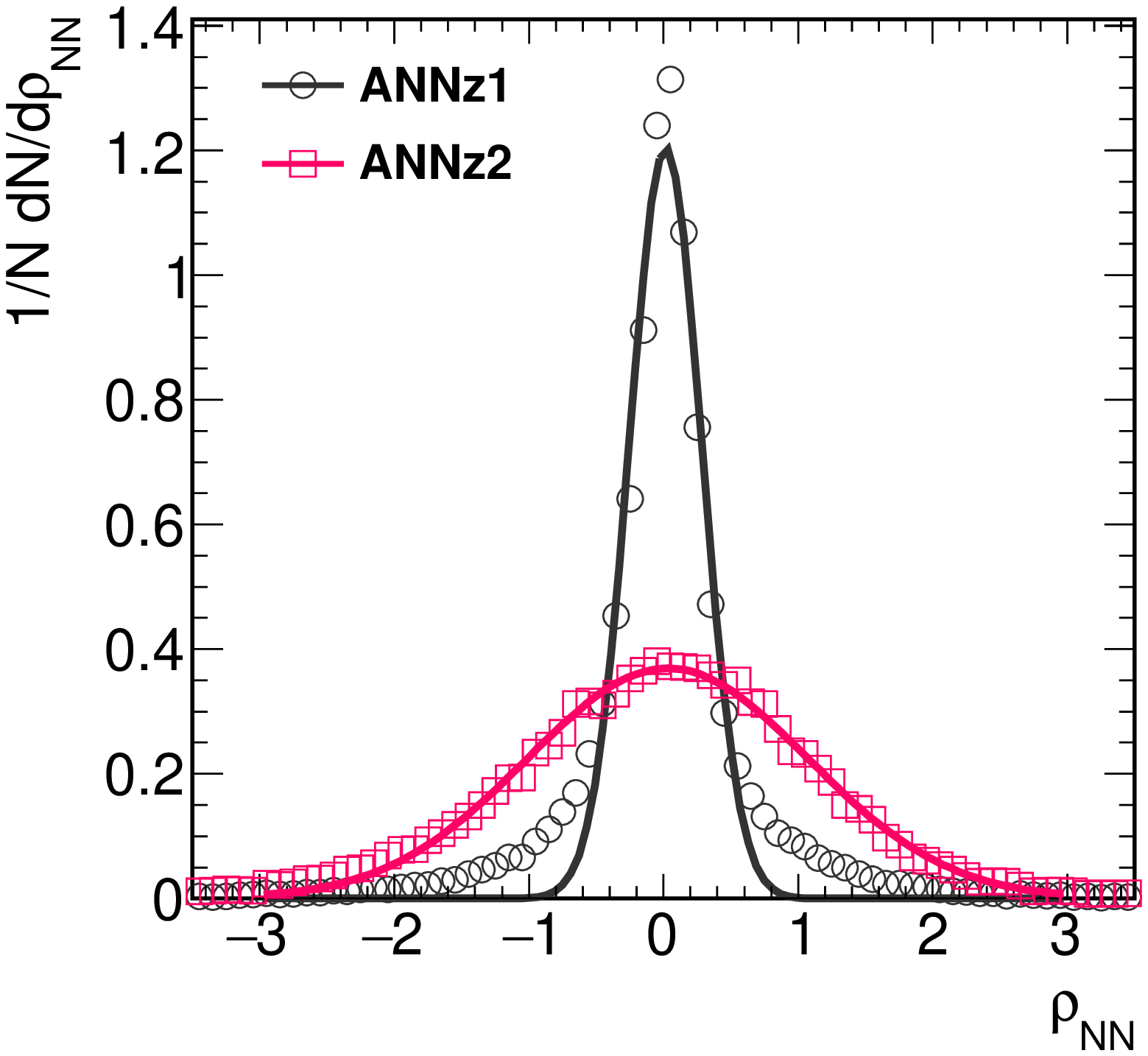}
  \caption{\label{FIG_boss_annz1_annz2_relZerr}
    Differential distributions of $\rho_{\mrm{NN}}$, the ratio between the \photoz bias
    and the associated uncertainty (see \autoref{knnRelErrorEq}), for the \photoz solutions
    derived using either \annzOne or \annz, as indicated. The markers represent the data
    and the lines represent fits to Gaussian functions. The fitted Gaussian width parameters
    are, respectively,~${0.27}$ and~${1.04}$ for \annzOne and \annz, where for
    well-representative uncertainty estimates, the expected value for the width is~$1$.
  }
\end{center}
\end{figure}

The distributions of $\rho_{\mrm{NN}}$ for our \annzOne and \annz solutions are shown in \autoref{FIG_boss_annz1_annz2_relZerr}.
We proceed by fitting a Gaussian function to each dataset.
We find that both distributions have a mean value which is
consistent with zero to a precision better than~$3\%$. In addition,
the distribution of $\rho_{\mrm{NN}}$ for \annzOne has a width
of~${0.27}$, while the corresponding value for \annz is~${1.04}$.
This indicates that the uncertainty estimation for
the \annz \photozs is significantly more reliable in comparison.

\subsection{Randomized regression PDF}\label{SECrandomRegression}
%
As mentioned above, we construct our PDF by combining multiple MLM estimators, each folded
with their respective single-value uncertainty estimator. The steps of the algorithm
may be summarized as follows:

\begin{enumerate}{}
\item
  A collection of MLMs is trained.
\item
  The ensemble of estimators goes through pre-selection, which includes ranking the solutions
  by their performance. The MLM which performs best is chosen as the single-value estimator.
\item
  The MLMs are folded with their corresponding intrinsic uncertainty, $\sigma_{\mathrm{gal}}$.
  They are then combined in different ways into a set of candidate-PDFs. The MLM combinations
  are chosen randomly, taking into account the ranking in performance.
\item
  The performance of the candidates is compared, using the parameter $\mathcal{C}$, defined below.
  The solution which best describes the true redshift distribution is selected as the final PDF.
\end{enumerate}

The first step in the calculation is the training of a set of 
\textit{randomized MLMs}. These differ from each other in several ways.
The latter includes setting unique random seed initializations, as well as changing the configuration
parameters of a given algorithm.\footnote{~See \autoref{SECtoyAnalysis}
and \autoref{SECappendix} for details.}
For instance, this may refer to using various types and numbers of neurons in an ANN,
or to arranging neurons in different layouts of hidden layers;
for BDTs, the number of trees and the type of boosting/bagging algorithm may be changed, etc.

In general, the choice of input parameters also has an effect on the performance~\citep{2015MNRAS.449.1275H}.
A randomized MLM therefore has the option to only use a subset of the given input parameters, or to train with predefined
functional combinations of parameters.
These combinations may also incorporate complicated scenarios. For instance, missing inputs for a specific object
may be mapped to predefined numerical values, such as the magnitude limits of the survey.

Additionally, \tmva provides the option to perform transformations on the
input parameters, including normalization and principal component decomposition. 
The transformations are done prior to training, as part of the internal pipeline of the code.
Applying transformations on inputs has the potential to improve the performance of
machine learning. For instance,~\cite{Soumagnac:2013odh} used principal component analysis
to augment their algorithm, by reducing the dimensionality of a classification task.
For \photoz inference, transformation are most useful when combining input observables
of different types, such as magnitudes and surface brightness.

Finally, the user may define training weights using functional expressions of both input parameters and observer
parameters (parameters not used directly for the training). The weights are applied during the training; they may
\eg be used to reduce the impact of noisy data on the result.
These may come in addition to the weights meant to account for unrepresentative training datasets,
which are discussed in the next section.

Once a set of randomized MLMs is initialized, the various methods are each trained.
Subsequently, a distribution of \photoz solutions for each galaxy is generated. 
A selection procedure is applied to the ensemble of answers, 
discarding outlier solutions which have very large values of $<\delta>$, $<\sigma_{68}>$
and $<f(2,3\sigma_{68})>$, compared to the entire ensemble.
The selected MLMs are then used to identify a single \photoz estimator, based on
the method with the best performance. The latter is denoted in the following as \zBest. 

In the next step, the various MLMs are 
folded with their respective single-value uncertainty. They are then
used in concert in order to
derive a complete probability distribution function.
The most trivial combination, is one in which we accept all MLMs with equal weights.
This, however, does not necessarily result in the best outcome, as the inclusion of
estimators with \eg large scatter, degrades the performance.
We therefore derive a dynamic weighting scheme for the combination of MLMs.
The weights are determined, using the cumulative distribution function (CDF) of a candidate-PDF,
\begin{equation}
  \mathcal{C}(\zSpec) = \int\limits_{z_{0} = 0}^{\zSpec} p_{\mrm{reg}}(z) \; \mrm{d}z \;.
\label{cumulDistEq}
\end{equation}
The latter is defined as the integrated PDF for redshifts smaller than some reference value, taken
here as the true redshift, \zSpec.
Here the differential PDF for a given redshift is denoted by $p_{\mrm{reg}}(z)$, and
$z_{0}$ corresponds to the lower bound of the PDF.

Let us consider a \photoz PDF which correctly describes the underlying redshift distribution. In this case,
one may think of \zSpec as a random variable which is distributed according to the PDF. It then follows that
$\mathcal{C}$ would be a flat distribution. As further illustration,
one may imagine the inverse problem. Supposing we generate a collection of random numbers,
uniformly distributed between~0 and~1. We then use these to
calculate $\mathcal{C}^{-1}$, the inverse of the CDF (the quantile function).
In this case, the distribution of $\mathcal{C}^{-1}$ values would
correspond to redshifts; it should then recover our PDF, assuming the PDF correctly
represents the underlying uncertainty on our \photoz inference.

The CDF of redshifts has previously been used to constrain \photoz PDFs,
as \eg in~\citet{MNR:MNR16765}. There it was the basis for
modifying PDFs which were constructed from likelihood functions, as part of a template fitting algorithm.
In \annz, $\mathcal{C}$ is used for the initial derivation procedure of the PDF.
This is done
by selecting from the collection of candidate-PDFs, the solution for which $\mathcal{C}$ is as close as possible
to a uniform distribution.

\subsection{Representativeness and completeness of the training sample}\label{SECrepresentativenessAndReliability}
%
Up to this point, we have discussed how the uncertainty on input parameters
and the differences between specific MLMs are treated in \annz. However,
machine learning methods based on training are susceptible to additional systematic effects.
Two possible sources of major bias come about for training datasets which are not
\textit{representative} or are not \textit{complete}.

One possible source of bias is the exact composition of the training dataset. Let us consider
an evaluated object from a photometric dataset, for which we have comparable training objects.
It is then important that the relative fraction of these training objects within the training sample be
the same as in the photometric dataset. If this is not
the case, the training sample is usually referred to as unrepresentative.

\begin{figure}[tb]
\begin{center}
  \includegraphics[trim=5mm 5mm 14mm 18mm,clip,width=.475\textwidth]{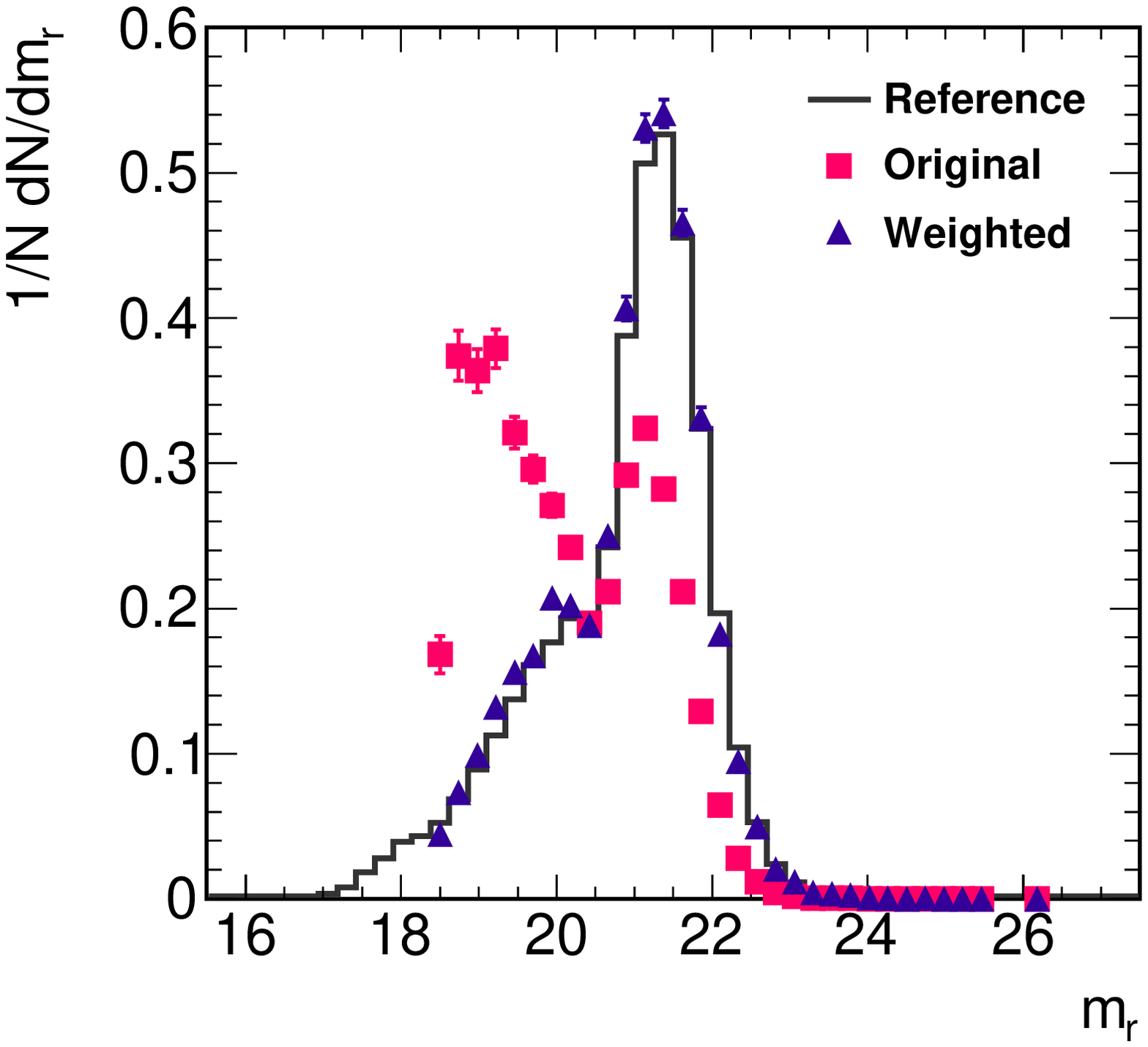}
  \caption{\label{FIG_boss_knnWgt}
    Differential distributions of the \ttt{r}-band magnitude, $\ttt{m}_{\ttt{r}}$, of objects in three samples, as indicated;
    the \textit{reference} sample, which corresponds to a hypothetical survey;
    the \textit{original} training sample, which is some spectroscopic dataset which is available for
    training an MLM;
    the \textit{weighted} training sample, which corresponds to the original training sample,
    after weights have been applied, as described in the text.
  }
\end{center}
\end{figure}

In order to illustrate the point, a simple example is shown in
\autoref{FIG_boss_knnWgt}. The figure includes the distributions of the \ttt{r}-band magnitude,
$\ttt{m}_{\ttt{r}}$, of objects in hypothetical
training and reference samples. The latter represents a complete and unbiased representation
of the $\ttt{m}_{\ttt{r}}$ of galaxies for some survey.
In this case, the distribution of $\ttt{m}_{\ttt{r}}$ in the training dataset is quite different from that in the
reference sample. An MLM trained using this training dataset will \eg give too high significance to training examples with
$\ttt{m}_{\ttt{r}}$ values close to~$19$.

The problem may be alleviated by reweighting the training sample. The purpose of the weights is to assign a correction
factor to galaxies as a function of the input parameters. The weighted distribution of galaxies should be such, that the
relative fraction of objects in each region in the parameter space is the same as in the reference sample. These weights are then
used as part of the training; they are also further propagated to the metric calculations,
to be used during the PDF optimization phase.
The reweighting procedure is implemented as part of the internal pipeline of \annz,
requiring only the definition of the reference dataset by the user of the code.

The weights are derived by matching the density of objects in the input parameter space
to that in the reference sample~\citep{Lima:2008kn}. This way, all inputs are reweighted simultaneously, accounting
for any intrinsic correlations. We derive the weights
using a \kdTree, calculating the number of neighbours of an object in
the training sample within some distance (see \autoref{knnDistanceEq}). We then find the number
of neighbours of the same object within the same distance, but in the reference sample. The weight is
finally taken as the ratio of these two numbers.

One may notice in \autoref{FIG_boss_knnWgt} that for ${\ttt{m}_{\ttt{r}} \lesssim 18.5}$, the
weighted training dataset does not match the reference sample.
The reason for this, is that the original training sample does not have any corresponding objects.
In this case, we usually refer to the training dataset as incomplete.
In general, an MLM should only be used on objects which have features that are represented in the training dataset.
In cases where no training examples exist, both the \photoz and the corresponding \photoz uncertainty are
equally unreliable.

\annz has a validation mechanism to check whether an evaluated object
falls under an incomplete region of the training sample.
Unfortunately, there is no systematic
way to correct the \photoz of objects which do not have comparable training examples.
These can instead be flagged as unreliable.

The algorithm uses a \kdTree to derive the density of objects from the training sample, which
have similar properties as the evaluated object.
We begin by computing ${R_{\mrm{NN}}^{\mrm{y/x}}}$,
the distance in parameter-space between the evaluated object, $x$, and
the closest corresponding object from the training sample, $y$ (see \autoref{knnDistanceEq}).
We then derive ${R_{\mrm{NN}}^{\mrm{y/n}}}$,
the distance from $y$, within which ${n_{\mrm{NN}}^{min}}$
objects from the training sample are found. Finally, we define our quality criteria as
\begin{equation}
  \qNN = \max \left\{\;0~,~\frac{ R_{\mrm{NN}}^{\mrm{y/n}} - R_{\mrm{NN}}^{\mrm{y/x}} }
                                      { R_{\mrm{NN}}^{\mrm{y/n}} } \; \right\} \;.
\label{qNNeq}
\end{equation}
\begin{figure}[p]
\begin{center}
    \subfloat[]{\label{FIG_boss_inTrainFlag_0}\includegraphics[trim=5mm 0mm 14mm 18mm,clip,width=0.475\textwidth]{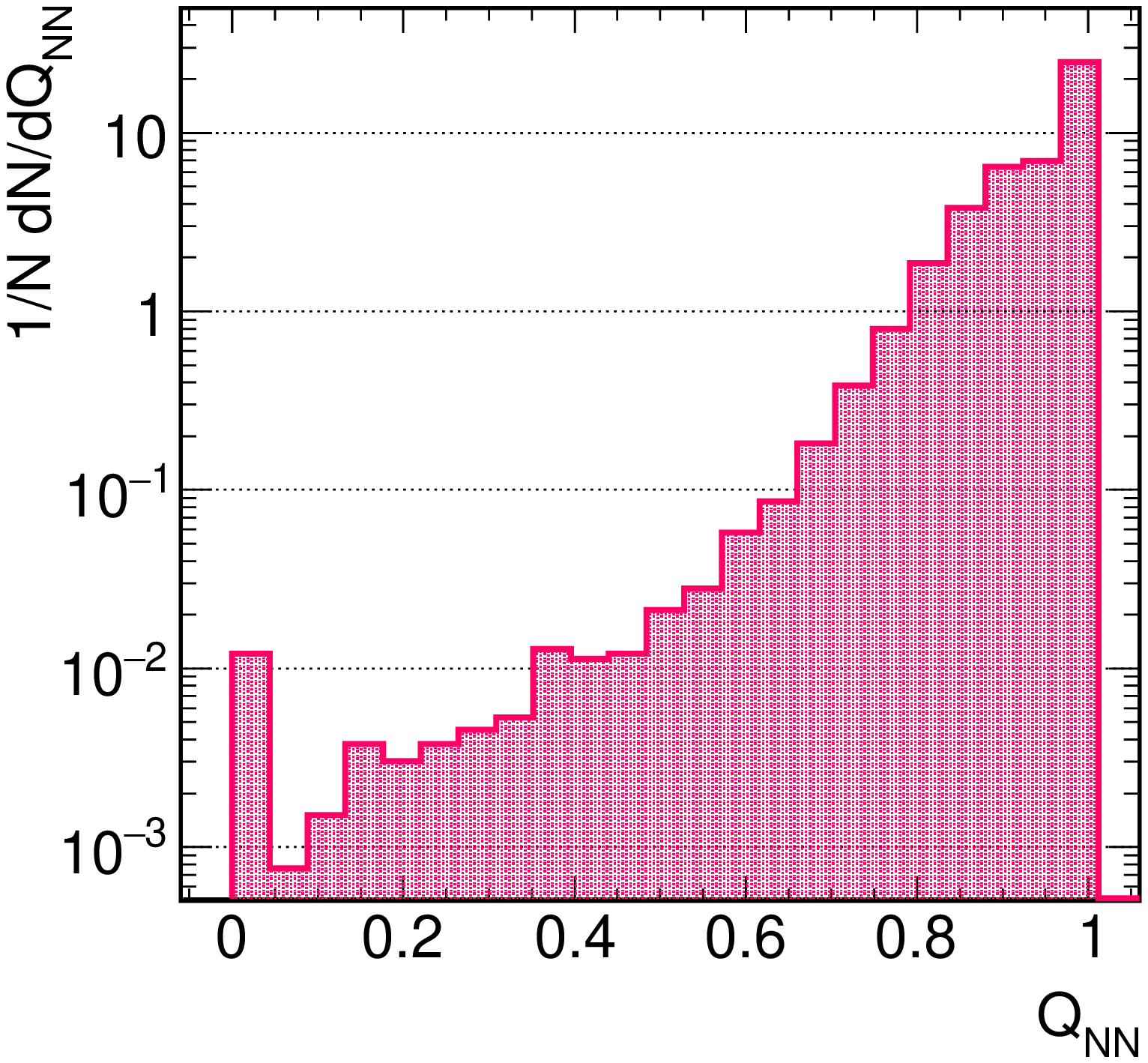}}
  \\
  %
    \subfloat[]{\label{FIG_boss_inTrainFlag_1}\includegraphics[trim=5mm 0mm 14mm 18mm,clip,width=0.475\textwidth]{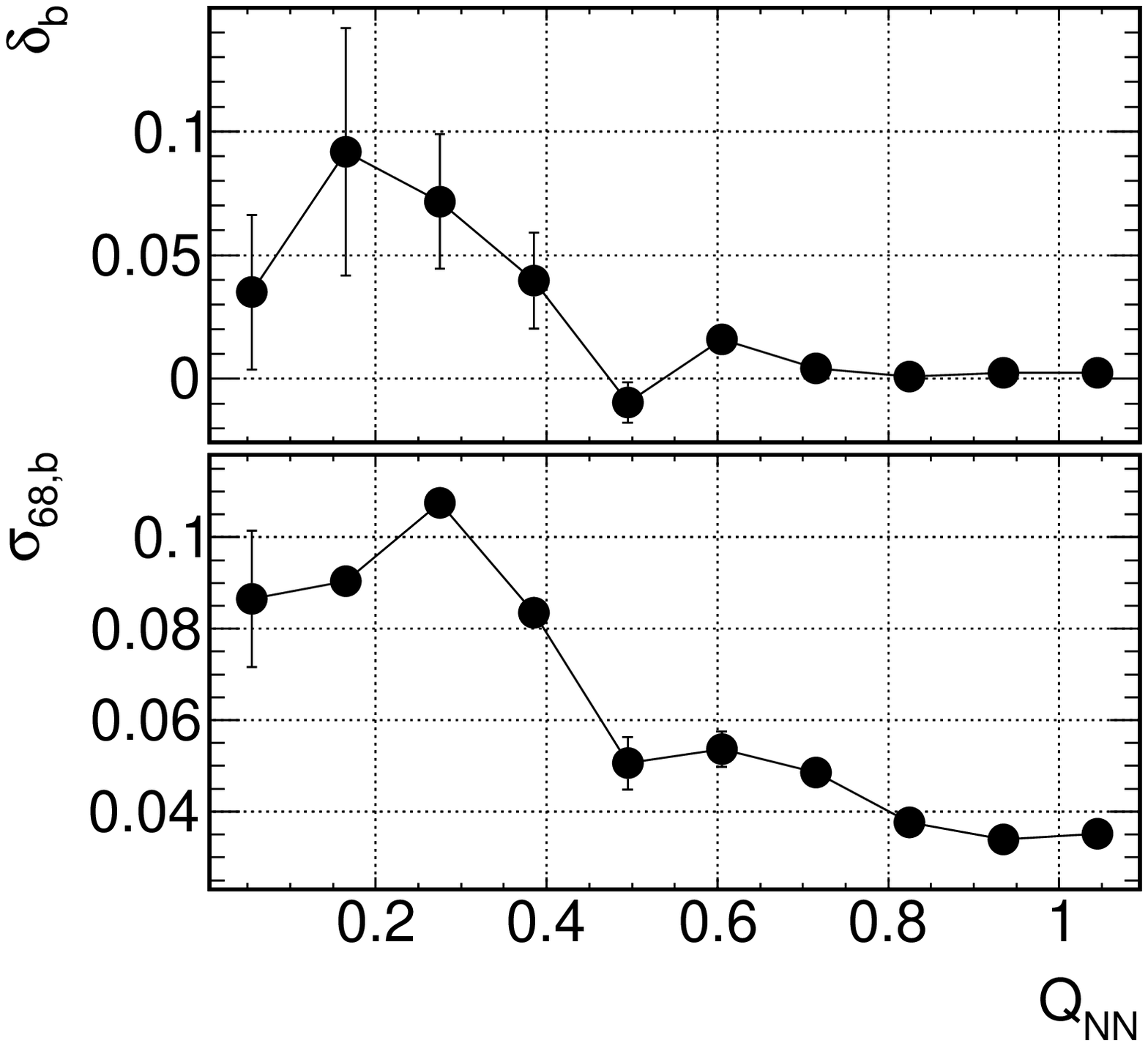}}
  %
  \caption{\label{FIG_boss_inTrainFlag}
    Properties of the quality criteria, \qNN (see \autoref{qNNeq}),
    for the hypothetical training and reference samples used for \autoref{FIG_boss_knnWgt},
    where the reference sample is taken as the evaluated dataset.
    \Subref{FIG_boss_inTrainFlag_0}~:~Differential distribution of \qNN.
    \Subref{FIG_boss_inTrainFlag_1}~:~Dependence of the \photoz bias,
    $\delta_{b}$, and of the \sixEightP scatter, $\sigma_{68,b}$, on \qNN.
  }
\end{center}
\end{figure} 

The parameter \qNN represents a typical distance-ratio between the evaluated object, and
similar training objects. For dense regions of the training sample, 
${R_{\mrm{NN}}^{\mrm{y/x}} \ll R_{\mrm{NN}}^{\mrm{y/n}}}$, which 
corresponds to ${\qNN \sim 1}$. Conversely, for sparse regions, one would have to search far
away in order to find object-$y$, resulting in low values of \qNN.
The steepness of the distribution of \qNN depends on the choice of ${n_{\mrm{NN}}^{min}}$,
and on the properties of the dataset.
We nominally use ${n_{\mrm{NN}}^{min} = 100}$, though this parameter
may be changed by the user of the code.

The parameter \qNN can be used to reject low-fidelity \photozs.
The exact cut on \qNN should be determined on a case-by-case
basis. It should take into account the fraction of excluded objects, and the
relative improvement in performance.
To illustrate the properties of \qNN, we use the hypothetical training and
reference samples defined for \autoref{FIG_boss_knnWgt}.
For the purpose of the example, we take the reference sample as the evaluated dataset.
The corresponding distribution of \qNN values is presented in \autoref{FIG_boss_inTrainFlag_0}.
We quantify our results in \autoref{FIG_boss_inTrainFlag_1}. Here,
we present the dependence on \qNN
of the \photoz bias, $\delta_{b}$, and of the \sixEightP scatter, $\sigma_{68,b}$.
As desired, the performance improves as the value of \qNN increases.
For this example, a conservative cut would be to reject galaxies with ${\qNN < 0.8}$.

\section{Performance of the estimators of \annz}
%
\subsection{Toy analysis}\label{SECtoyAnalysis}
\begin{figure}[tb]
\begin{center}
  \includegraphics[trim=5mm 5mm 14mm 18mm,clip,width=.475\textwidth]{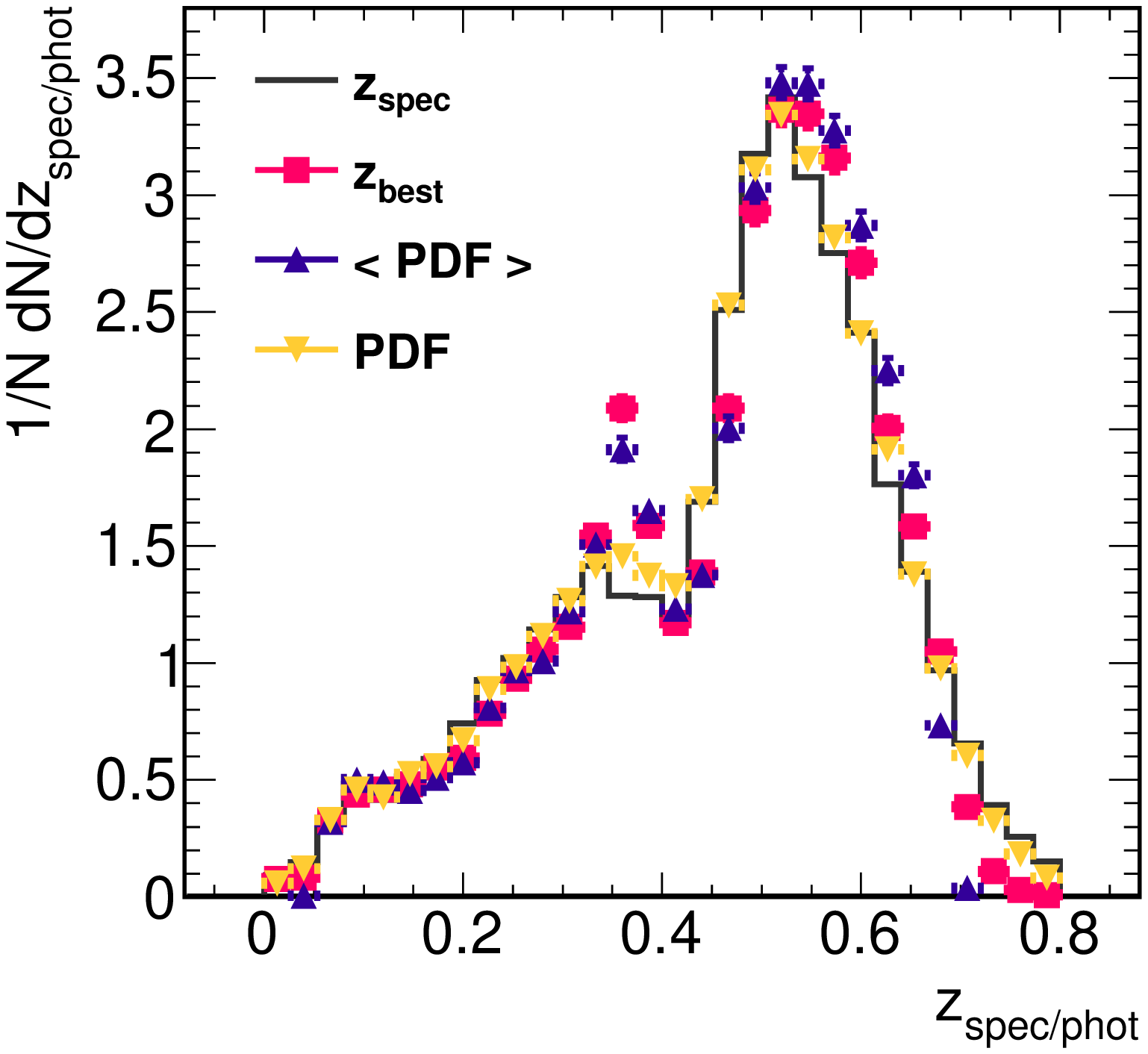}
  \caption{\label{boss_NzNominal}
    Differential distributions of the spectroscopic redshift, \zSpec,
    and of the respective photometric redshift, \zPhot, 
    where \zBest is the single-value MLM solution with the best performance, $< \PDF >$ is the single-value average
    of the PDF solution, and \PDF is the full stacked PDF, as indicated. The overall fit of the stacked PDF to the true
    redshift distribution is better than that of the single-value solutions.
  }
\end{center}
\end{figure}

\begin{figure*}[tbp]
\begin{center}
  \begin{minipage}[c]{0.32\textwidth}
    \includegraphics[trim=3mm 60mm 20mm 60mm,clip,width=1.\textwidth]{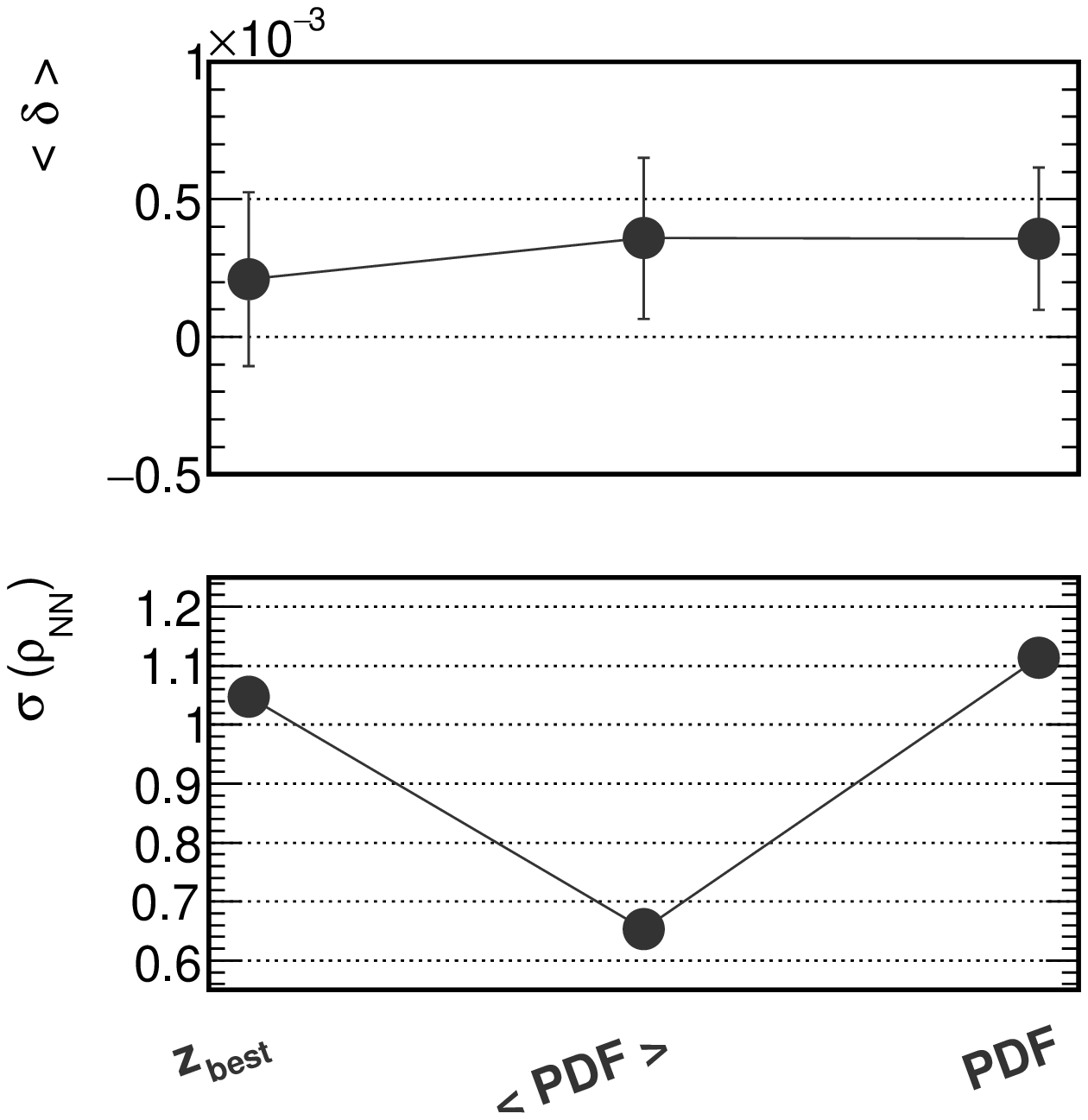}
  \end{minipage}\hfill ~
  \begin{minipage}[c]{0.32\textwidth}
    \includegraphics[trim=3mm 60mm 20mm 60mm,clip,width=1.\textwidth]{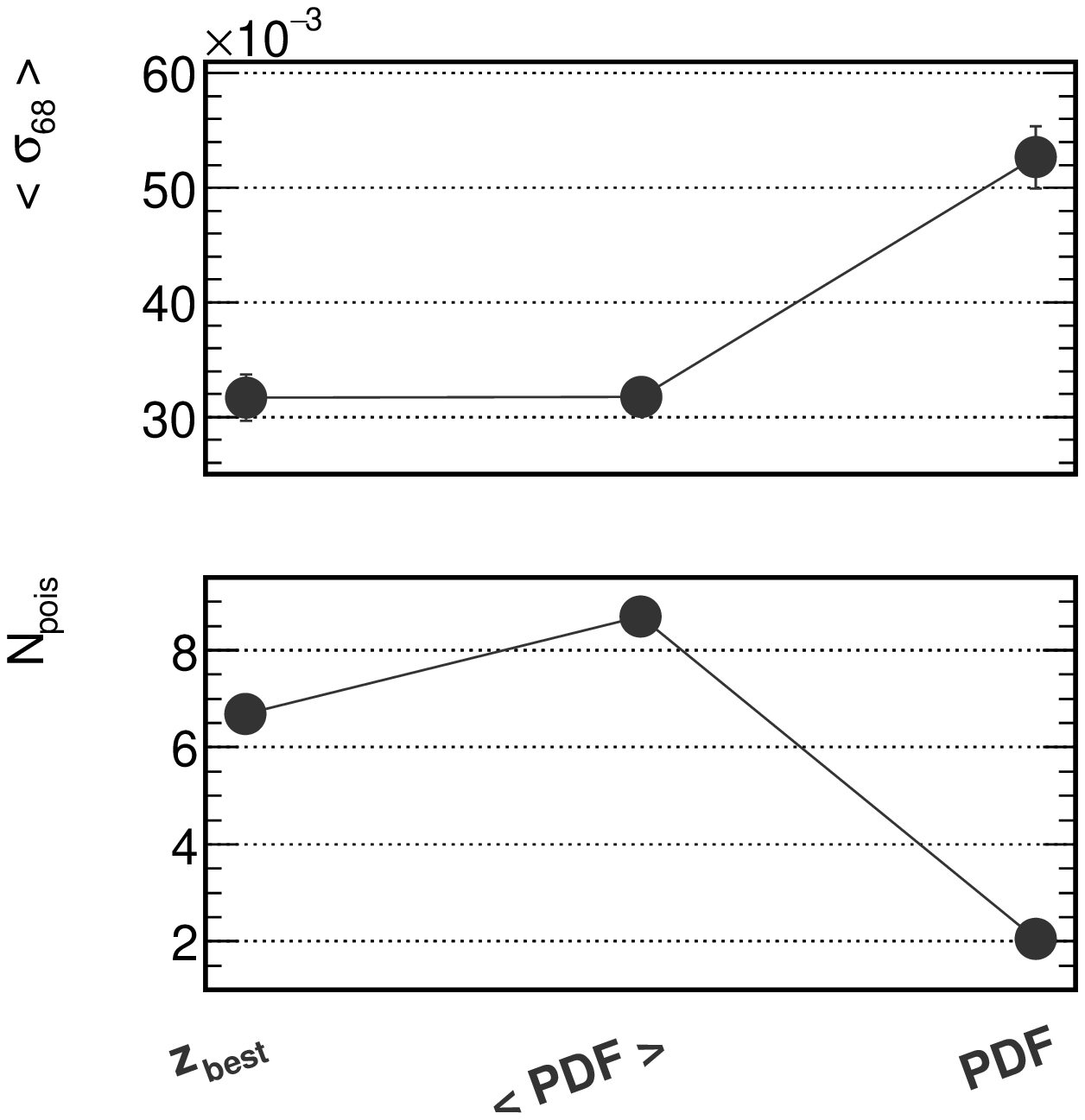}
  \end{minipage}\hfill ~
  \begin{minipage}[c]{0.32\textwidth}
    \includegraphics[trim=3mm 60mm 20mm 60mm,clip,width=1.\textwidth]{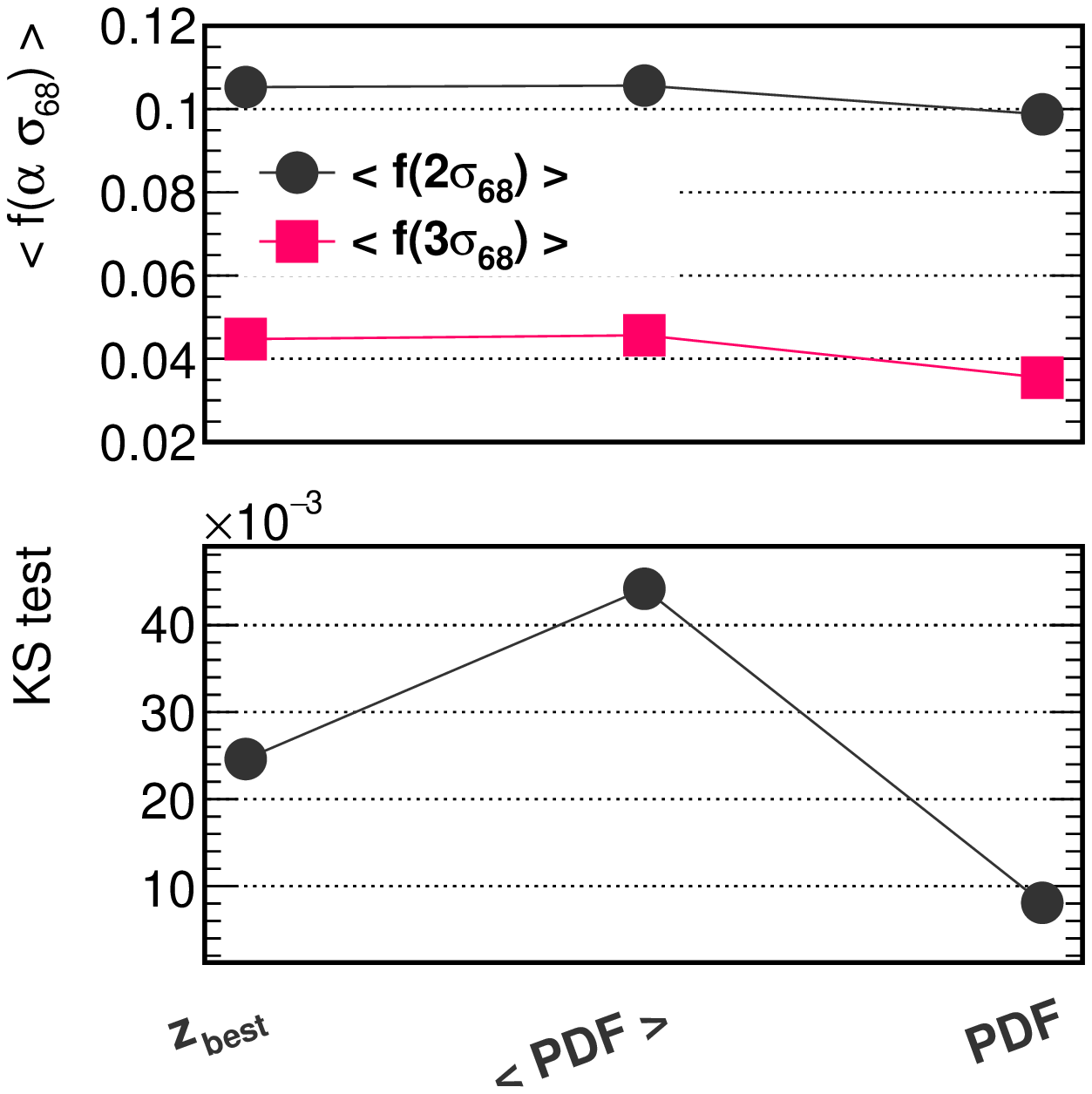}
  \end{minipage}\hfill
  \caption{\label{boss_metricsNominal}
    \Photoz metrics, averaged over the entire redshift range,
    for the nominal solutions of \annz, 
    where \zBest is the single-value MLM solution with the best performance, $< \PDF >$ is the single-value average
    of the PDF solution, and \PDF is the full stacked PDF, as indicated. The metrics are
    the bias, $<\delta>$; the \sixEightP scatter, $<\sigma_{68}>$;
    the outlier fractions, $f(\alpha\sigma_{68})$, for~${\alpha=2\;\mrm{or}\;3}$;
    the $N_{\mrm{pois}}$ and $\mrm{KS}$-test statistics;
    and $\sigma(\rho_{\mrm{NN}})$, the \sixEightP width of the distribution of $\rho_{\mrm{NN}}$ (see \autoref{knnRelErrorEq}).
    The lines are meant to guide the eye.
    All three solutions have comparable values of \photoz bias. The stacked PDF solution exhibits a relatively larger
    scatter, due to the inclusion of the tails the distribution in the calculation. However, the overall fit
    of the full PDF to the true redshift distribution, indicated by $N_{\mrm{pois}}$ and the $\mrm{KS}$-test, is
    better in comparison.
  }
\end{center}
\end{figure*}

%
\Autoref{boss_NzNominal} shows the distribution of the nominal \photoz estimators of \annz for our SDSS dataset.
These include the single-value \photoz estimator, \zBest,
the single-value average of the randomized regression PDF, $< \PDF >$, and
the full PDF solution, \PDF.
The corresponding performance metrics are presented in \autoref{boss_metricsNominal};
the bias, $<\delta>$; the \sixEightP scatter, $<\sigma_{68}>$;
and the outlier fractions, $<f(2\sigma_{68})>$ and $<f(3\sigma_{68})>$.
In addition, we include the metric $\sigma(\rho_{\mrm{NN}})$, defined as the \sixEightP width of the
distribution of $\rho_{\mrm{NN}}$ (see \autoref{knnRelErrorEq}).
Finally, the $N_{\mrm{pois}}$ and $\mrm{KS}$-test statistics of the various $N(\zPhot)$ distributions are
shown as well.

The \zBest solution is the same one shown in~\autoref{FIG_boss_singleANNmetrics}, and corresponds to
an ANN with architecture $\{N,N+1,N+9,N+4,1\}$, where $N$ corresponds to the
number of input parameters (in this case, five magnitudes).
The ensemble of MLMs used for the PDF is composed of 50~ANNs and 50~BDTs, with specific MLM
options chosen at random as described next. In addition, for both the ANNs and the BDTs, the
input parameters for the training were chosen as either the five magnitudes, combinations of
magnitude and colours, or subsets of the latter. Furthermore,
variable transformations on the input parameters (normalization, principal component analysis, decorrelation)
were switched on or off at random.

The ANNs were configured with variations of the following
parameters:\footnote{~See \autoref{SECappendix} for details on the configuration options.}
the numbers of hidden layers was varied between~2 and~4;
the number of neurons in a hidden layer was varied between~$N$ and~${(N+10)}$;
the neuron activation function was chosen as either a \ttt{sigmoid} or a \ttt{tanh} function;
use of a regulator was switched on or off;
the number of steps between convergence tests was randomized between~100 and 500~steps;
the MLMs were trained using back-propagation, with or without the use of second derivatives
of the ANN error function.

The BDTs were defined using the following settings:
the number of trees was randomized between~300 and ~1200;
the boosting algorithm was changed between the available options in \tmva;
the threshold criteria for splitting nodes was varied between~$0.1\%$ and~$1\%$ of the
number of training objects per node;
the separation criteria for testing node-splitting was chosen at random.

We observe that the three \photoz estimators, \zBest, $< \PDF >$ and \PDF, all have
an average bias which is consistent with zero. Comparing the scatter,
the full PDF has a larger scatter relative to the single-value estimators.
This is expected, as the calculation for PDFs is performed bin-by-bin, taking
into account the tails of the PDF.
For approximately symmetric PDFs, the tails cancel out. They therefore
do not affect the bias or scatter of the average of the PDF.
However, for the full solution, the negative and
positive contributions from the tails increase $<\sigma_{68}>$.
The scatter for the full PDF is therefore larger by construction.
The increased value of the PDF scatter is not a disadvantage. Rather, it
represents a more realistic estimation of the underlying uncertainty on the \photozs.
This is reflected by the value of $\sigma(\rho_{\mrm{NN}})$,
which is much better (closer to~1) for the
\PDF estimator, than for its average.

Finally, the shape of the full PDF provides a better description
of the underlying redshift distribution, as expressed
by the low values of the $N_{\mrm{pois}}$ and $\mrm{KS}$-test statistics.
The difference in the performance may even be appreciated by eye
from \autoref{boss_NzNominal}. Specifically, the stacked \PDF
provides a better estimation of the true redshift for
${\zSpec\sim0.35}$ and for ${\zSpec>0.7}$, where the single-value
solutions are less precise.
%
%

\subsection{Other applications of \annz}
%
We have only commented here on performance metrics for \annz, such as \photoz bias,
and compatibility with the underlying redshift distribution. However,
to fully qualify the algorithm, one would need to perform a Cosmological
analysis involving photometric redshifts~\citep{Rau:2015sha}.
Such a study is beyond the scope of the current work. However,
\annz has already been used for several DES analyses, and is included
in the first public data release of the experiment.

In~\cite{2015arXiv150705909B}, the performance of \annz
was compared with that of three other codes, \skynet, \tpz and \bpz. The first two are machine-learning codes which
employ a similar algorithm using different MLM types, while \bpz is a template-fitting code. The comparison was done
in the context of the first DES Cosmology results~\citep{2015arXiv150705552T}, where the difference between the \photozs
were propagated to the systematic uncertainty on a weak lensing analysis. 
One should notice that~\cite{2015arXiv150705909B} performed the comparison between the
different estimators by first assigning galaxies to one set of redshift bins. The latter were
determined by the nominal code in the study, \skynet. The \photozs
of the various codes for a given galaxy sub-sample (\skynet \photoz bin) were then compared.
This may produce a selection bias; effectively, the PDF for each code is constrained by the results of \skynet.
However, even given this possible bias, \annz was found to be compatible with the other codes.

In another study~\citep{2015arXiv150705647L}, a systematic test of variations in the observing conditions in DES was performed,
comparing \annz, \tpz and \bpz. In this case, it was shown that \annz minimizes the variations in the \photoz distribution
due to degraded input data, and that it reduces the amount of outliers.

\section{Summary}
%
\annz is a new major version of the public photometric redshift estimation software, first
developed by~\citet{Collister:2003cz}. It has already been used as part of the first
weak lensing analysis of the Dark Energy Survey, and is included in the first data release of
the experiment. The code is also planned to be incorporated in
the software pipelines of future projects.
In this paper we have introduced the algorithm available in the new implementation,
and have illustrated the performance of the code using spectroscopic data. 

\annz incorporates several machine learning methods, such as artificial neural networks and boosted
decision/regression trees.
The different algorithms are used in concert in order to optimize the \photoz reconstruction,
and to estimate the associated uncertainties.
This is done by generating a wide selection of machine learning methods, utilizing \eg different
ANN architectures and BDT algorithms. The final product of \annz is either
a single-value \photoz estimator, or a full \photoz probability distribution
function. PDF derivation is an important new feature of \annz, not available in the previous
version of the code.

PDFs are calculated by \annz using two different approaches.
The nominal approach is a new technique, called randomized regression.
In this mode, optimization is performed by ranking the
different solutions according to their performance, which is
determined by the respective \photoz bias, scatter and outlier fraction parameters.
The single solution with the best performance is chosen as the nominal \photoz estimator
of \annz. In addition, the entire collection of solutions is used in order to derive
a \photoz PDF. The PDF is constructed in two phases.
In the first phase, each solution is folded with a distribution of uncertainty values,
which is derived using the KNN uncertainty estimation method.
In the second phase, the ensemble of solutions is combined. This is done using
dynamically determined weighting schemes, intended to optimize the final PDF.
Additionally, we have implemented in \annz a second approach for PDF-derivation,
called binned classification. The latter
has been used in the past, and is not discussed in the current paper.

\annz also includes an implementation of a method to correct for training samples which are not representative of
the features of the evaluated dataset. In addition, we introduce a new
method to account for samples which are not complete.
The former is performed by applying weights to training objects during training and during
\photoz optimization, in order to match the properties of the evaluated dataset.
For the latter, a quality flag is generated for each evaluated object. The flag indicates
whether the derived \photoz solution is reliable, based on the completeness of the sample.

\section*{Acknowledgements}
%
We would like to thank 
Manda Banerji,
Christopher Bonnett,
Antonella Palmese and
Maayane Soumagnac
for the useful discussions regarding the nature of photometric redshifts and machine learning.
We would also like to thank the \photoz working groups of DES and of the
Euclid experiment for giving feedback on the code.

OL acknowledges an European Research Council Advanced Grant FP7/291329, which also supported IS.
FBA acknowledges the Royal Society for a Royal Society University Research Fellowship.

This work uses publicly available data from the SDSS.
Funding for SDSS-III has been provided by the Alfred P.~Sloan Foundation,
the Participating Institutions, the National Science Foundation, and the U.S.\
Department of Energy Office of Science. The SDSS-III website
is \href{http://www.sdss3.org/}{http://www.sdss3.org/}.

\bibliographystyle{astron} \bibliography{bib}

\appendix
\section*{Appendix}
\section{Quick-start guide}\label{SECappendix}

To illustrate the use of \annz, we provide a short guide for running the code.
The following is limited to describing the randomized regression mode,
corresponding to version $\ttt{2.1.2}$ of the code.
Please see the on-line
documentation\footnote{~See \href{https://github.com/IftachSadeh/ANNZ}{https://github.com/IftachSadeh/ANNZ}~.}
for further details, as well as up to date instructions.

\subsection{Work-flow}

Randomized regression is run using the following consecutive shell commands. In this
example, the commands employ the \ttt{Python} control script,
$\ttt{annz\_rndReg\_quick.py}$, which is provided with the code:
\lstset{language=Python,caption={}} \begin{lstlisting}
python scripts/annz_rndReg_quick.py --randomRegression --genInputTrees
python scripts/annz_rndReg_quick.py --randomRegression --train
python scripts/annz_rndReg_quick.py --randomRegression --optimize
python scripts/annz_rndReg_quick.py --randomRegression --evaluate
\end{lstlisting}
These correspond to the four stages of the pipeline: data processing, training, optimization, and evaluation.

In the following, we describe each of these stages. We use \ttt{Python} pseudo-code to
represent the content of the example script. The dictionary syntax, \lstinline{ANNz["XXX"]},
stands for a job-option parameter labelled \lstinline{XXX}, which is exposed to \annz. All other
variables are internal to the control script.

\subsection{Data processing}
In the initial stage, the training and validation samples defined by the user are ingested. If the user
does not explicitly define separate input files for training and for validation, the complete sample
is randomly split.

The user also has the option to define a reference sample, which represents
the dataset which is eventually evaluated. If this reference is provided, training weights are calculated,
as described in \autoref{SECrepresentativenessAndReliability}.

For example, the user may define,
\lstset{language=Python,caption={},breaklines=no} \begin{lstlisting}
ANNz["inAsciiFiles"]          = "trainingTestingSample.csv"
ANNz["inAsciiVars"]           = "F:m_u ; F:e_u ; F:m_g ; F:e_g ; D:z_spec ; C:survey"

ANNz["useWgtKNN"]             = True
ANNz["inAsciiFiles_wgtKNN"]   = "referenceSample.csv"
ANNz["inAsciiVars_wgtKNN"]    = "F:m_u ; F:e_u ; F:m_g ; F:e_g"
ANNz["weightVarNames_wgtKNN"] = "m_u ; m_g ; e_u; e_g ; (m_u-m_g)"
\end{lstlisting}

Here \lstinline{inAsciiFiles} defines the input file containing
the dataset for training and validation. The corresponding list of variables in this file
is defined in \lstinline{inAsciiVars}. For brevity, we define only a few inputs here; these are formatted as
a semicolon-separated list of variable type and name. The former are \eg \lstinline{F}, standing for floating precision,
\lstinline{D}, standing for double precision, and \lstinline{C}, standing for a string variable.
The variable names, \lstinline{m_u}, \lstinline{e_u}, \lstinline{m_g} and \lstinline{e_g}
stand in this example for a pair of magnitudes and their corresponding errors; the variable \lstinline{z_spec} stands for the
spectroscopic redshift; the variable \lstinline{survey} stands for the name of the spectroscopic survey.
We note that our use of magnitudes, while useful for \photoz estimation, has no particular significance.
The user may assign any type of input (with any assigned name) as part of the input dataset.

Setting the variable \lstinline{useWgtKNN} to \lstinline{True} activates the calculation of training weights. The associated parameters are
\lstinline{inAsciiFiles_wgtKNN} and \lstinline{inAsciiVars_wgtKNN}, 
respectively used to define the file-name of the reference sample,
and the corresponding list of variables it contains. The parameter, \lstinline{weightVarNames_wgtKNN} defines the variables which
are used for the KNN search. In this example, distance between neighbours
is defined in magnitude (\lstinline{m_u}, \lstinline{m_g}), in magnitude-error
(\lstinline{e_u}, \lstinline{e_g}) and in colour (\lstinline{m_u-m_g}). Any functional combination of input
parameters may be used for the KNN search, for any variable which is defined in both \lstinline{inAsciiVars}
and \lstinline{inAsciiVars_wgtKNN}.

Once training weights are calculated, they are propagated automatically to all calculations in the following
stages. This includes the training of MLMs, the optimization process and the performance plots generated
as part of the output of the code.
The training weights themselves are also included as part of the output of \annz, for
every object from the training and validation samples. \annz may thus also be used
to calculate representativeness weights for use by other codes.

\subsection{Training}
In the second stage of the pipeline, a collection of MLMs is trained. The MLMs may
be trained consecutively, or in parallel (\eg using a batch-system).

For example, the user may set the following options:
\lstset{language=Python,caption={},breaklines=no} \begin{lstlisting}
ANNz["zTrg"]    = "z_spec"
ANNz["minValZ"] = 0.0
ANNz["maxValZ"] = 0.8
ANNz["nMLMs"]   = 20

for id in range(ANNz["nMLMs"]): 
  if   (id % 3) == 0: vars = "m_u ; m_g"
  elif (id % 3) == 1: vars = "m_u ; m_g ; (m_u-m_g) ; e_u ; e_g"
  else:               vars = "(m_u*(m_u < 25) + 25*(m_u >= 25))
                              ; (m_g*(m_g < 23.5) + 23.5*(m_g >= 23.5))"
  ANNz["inputVariables"] = vars 

  if   id == 0: opt = "ANNZ_MLM=ANN : HiddenLayers=N,N+5 : NeuronType=sigmoid
                       : UseRegulator=True : TrainingMethod=BFGS : NCycles=500"
  elif id == 1: opt = "ANNZ_MLM=BDT : NTrees=600 : MinNodeSize=2%
                       : BoostType=AdaBoost : VarTransform=N,D,P"
  elif id == 2: opt = "ANNZ_MLM=KNN : nkNN=90"
  else:         opt = ""
  ANNz["userMLMopts"] = opt

  ANNz["userCuts_train"]    = "(e_u < 5) && (survey == \"SDSS\")"
  ANNz["userCuts_valid"]    = "e_u < 10"
  ANNz["userWeights_train"] = "1/((1+e_u)*(1+e_g))"
  ANNz["userWeights_valid"] = ""
\end{lstlisting}

The target of the regression (the spectroscopic redshift) is defined in
\lstinline{zTrg}, with the allowed limits for the latter
set in \lstinline{minValZ} and \lstinline{maxValZ}.
In this case, $20$~randomized MLMs will be trained (specified by \lstinline{nMLMs}).
The variables used as input for the training are defined in \lstinline{inputVariables}. One
can select any functional combination of the available parameters which have
previously been defined in \lstinline{inAsciiVars}, including logical expressions.
An example for the latter is the choice made for the third option.
Here magnitudes are mapped to some effective magnitude-limit,
which may prevent training with noisy data.

The type of MLM for each of the randomized ensemble is defined in the \lstinline{userMLMopts}
parameter. The current example shows configurations of an ANN, a BDT and a KNN.
Here, the ANN is defined as having two hidden layers, the first with \lstinline{N}
and the second with \lstinline{N+5} neurons, where \lstinline{N} is the number of
input parameters; the selected type of neuron is a \lstinline{sigmoid} function;
a regulator is used for the training; the training method is chosen as \lstinline{BFGS}
(using second derivatives of the error function);
a maximum of $500$~training cycles are allowed.
The BDT is defined as being composed of $600$~trees (\lstinline{NTrees}); a minimum of
$2\%$~of training objects is included in each tree-node (\lstinline{MinNodeSize});
training employs the \lstinline{AdaBoost} boosting algorithm (\lstinline{BoostType}).
The KNN in this example is defined simply as using $90$~near neighbours.

Of the key-words defined for \lstinline{userMLMopts}, the only pattern unique to \annz is \lstinline{(ANNZ_MLM = XXX)},
here with \lstinline{XXX} being \lstinline{ANN}, \lstinline{BDT} or \lstinline{KNN}. This tag defines for
\annz which MLM type to use. All other job-options are native to \tmva.
For instance, an ANN may be trained with \lstinline{TrainingMethod = BP}, \lstinline{GA},
or \lstinline{BFGS}; a BDT may use boosting
(\lstinline{BoostType = AdaBoost}, \lstinline{RealAdaBoost},
\lstinline{AdaBoostR2}, or \lstinline{Grad}), or it may use bagging (\lstinline{ABaggingNN}), etc.
The various possible settings are defined in the \tmva
manual,\footnote{~See
\href{http://tmva.sourceforge.net/optionRef.html}{http://tmva.sourceforge.net/optionRef.html}~.}
along with overviews of the corresponding algorithms.
All machine learning methods available through \tmva may be used in \annz. However,
in our experience, ANNs and BDTs perform best for the task of \photoz inference.

In the current example, the user has also requested that
the variables used for training the BDT will have gone through transformations
prior to training. The latter are defined using the \lstinline{VarTransform}
parameter, with \lstinline{N} representing normalization, \lstinline{D},
decorrelation, and \lstinline{P} standing
for principle-component decomposition. The \lstinline{VarTransform} flag
may be added to any \lstinline{userMLMopts} option string, for any type of MLM.
The transformations are performed as part of the internal pipeline of the code,
and are automatically applied to evaluated objects.

The empty selection for \lstinline{userMLMopts} indicates for \annz
that MLM configuration parameters should be chosen on the fly. This is
done as part of the internal pipeline of the code, and results in randomized configurations
of ANNs and BDTs.

In the example, we also show how the user may define cuts for the training and validation
samples (\lstinline{userCuts_train}, \lstinline{userCuts_valid}).
For instance, assuming we have spectroscopic data from several surveys, the user has chosen
to only train with galaxies from the \lstinline{"SDSS"} survey. In addition, a cut is set to only use
objects with \lstinline{e_u} below certain limits. Such choices are useful for comparing the performance
for different training sub-samples. 
Additionally, weights may be defined for the training and validation samples using
\lstinline{userWeights_train} and \lstinline{userWeights_valid}. These take effect in addition
to the representativeness weights, provided the latter were calculated in the previous stage.
We note that the various cut and weight expressions can be set to different values 
for each of the randomized MLMs. For instance, the user may choose to impose a cut on
magnitude errors for half of the randomized MLMs, to asses if such a constraint improves
the performance or not.

\subsection{Optimization}
In the optimization stage, the performance of the ensemble of trained MLMs is derived.
The optimal solution is chosen as \zBest, and a PDF is derived.

There are several control options which the user may set,
\lstset{language=Python,caption={},breaklines=no} \begin{lstlisting}
pdfBinsType = 0
if   pdfBinsType == 0: ANNz["userPdfBins"] = "0.0 ; 0.2 ; 0.3 ; 0.4 ; 0.5 ; 0.6 ; 0.8"
elif pdfBinsType == 1: ANNz["nPDFbins"]    = 90
elif pdfBinsType == 2: ANNz["pdfBinWidth"] = 0.01

ANNz["max_bias_PDF"]    = 0.01
ANNz["max_sigma68_PDF"] = 0.044
ANNz["max_frac68_PDF"]  = 0.10
ANNz["MLMsToStore"]     = "LIST ; 0 ; 1 ; 3"
\end{lstlisting}

The first block shows how the user may define the binning-scheme for the PDFs. One may set
one of the following: \lstinline{userPdfBins} can be used to define a specific set
of bins; \lstinline{nPDFbins} can be used to divide the allowed range of the regression target
into (in this case) $90$~bins of equal width; \lstinline{pdfBinWidth} can be used to
divide the allowed range into a dynamically determined number of bins, which all
have a width of (in this case) ${0.01}$.

In general, all derived MLMs are combined to form the PDF. However, it is possible
to set exclusion criteria, and reject those which perform badly.
The parameters \lstinline{max_bias_PDF}, \lstinline{max_sigma68_PDF} and \lstinline{max_frac68_PDF}
represent these criteria; these respectively define
upper limits on the values of the bias, the \sixEightP scatter, and corresponding combined outlier
fraction. Individual MLMs with metric values higher than the upper limits, are not incorporated into the PDF.

By default, only \zBest and the PDF are included in the output of \annz.
However, it is possible for the user to define additional MLM estimators to be written out.
This is done using the \lstinline{MLMsToStore} parameter, which may include any MLM-\lstinline{id}
in the range, ${0 \leq \lstinline{id} < \lstinline{nMLMs}}$.

\subsection{Evaluation}
In the final stage of the pipeline, the user defines a dataset, for which the \photoz estimators are calculated.
Additionally, the quality parameter for incomplete training, \qNN, can be calculated on request.

For example, it is possible to choose the following configuration:
\lstset{language=Python,caption={},breaklines=no} \begin{lstlisting}
ANNz["inAsciiFiles"]           = "evaluatedSample.csv"
ANNz["inAsciiVars"]            = "F:m_g ; F:e_u ; F:m_u ; F:e_g"

ANNz["addInTrainFlag"]         = True
ANNz["weightVarNames_inTrain"] = "m_u ; m_g ; (m_u-m_g)"
ANNz["minNobjInVol_inTrain"]   = 150
\end{lstlisting}
where the \lstinline{inAsciiFiles} and \lstinline{inAsciiVars} variables are set as
for the initial data processing stage. We note that \lstinline{inAsciiVars} does not need to
exactly correspond to the same structure as for the previous stages. However, it must include
all variables which were used for training MLMs (see \lstinline{inputVariables}).

If the \lstinline{addInTrainFlag} parameter is set to \lstinline{True}, the
\qNN estimator is added to the output. For the calculation of \qNN, the user needs
to define \lstinline{weightVarNames_inTrain}, the list of
variables to be used for the KNN search. The user also has the option to
set the value of ${n_{\mrm{NN}}^{min}}$ (see \autoref{SECrepresentativenessAndReliability}), using
the parameter, \lstinline{minNobjInVol_inTrain}.

\end{document}